\documentclass[12pt]{article}
\pdfoutput=1

\usepackage{amsmath,amsfonts,amssymb,epsfig}
\usepackage{slashed}
\numberwithin{equation}{section}

\usepackage{relsize}

\usepackage{hyperref}

\usepackage{xcolor}
\hypersetup{
    colorlinks,
    linkcolor={red!0!black},
    citecolor={blue!50!black},
    urlcolor={blue!80!black}
}
\definecolor{dblue}{rgb}{0,0,0.8}
\usepackage{xcolor}
\hypersetup{
    colorlinks,
    linkcolor={red!0!black},
    citecolor={blue!50!black},
    urlcolor={blue!80!black}
}
\definecolor{dblue}{rgb}{0,0,0.8}
\definecolor{orange}{rgb}{0.8,0.4,0}
\definecolor{darkred}{rgb}{0.5,0.,0}

\usepackage{graphicx}
\usepackage{moresize}
\usepackage{cancel}
\usepackage{cite}
\definecolor{orange}{rgb}{0.8,0.4,0}
\definecolor{darkred}{rgb}{0.5,0.,0}

\usepackage{graphicx}
\usepackage{moresize}
\usepackage{cancel}
\usepackage{cite}
\usepackage{color}
\usepackage{multirow}

\usepackage{xspace}
\usepackage[width=1\textwidth,margin={30pt,0pt},font=footnotesize,labelfont=bf]{caption}
\usepackage{enumitem}

\def\s{\sigma}

\def\d{\partial}

\def\nn{\nonumber}
\def\Rb{R_b}
\def\dc{\Delta c}

\newcommand{\scr}[1]{\ensuremath{\mathcal{#1}}}

\newcommand{\ep}{\ensuremath{\epsilon}}

\def\ep{\epsilon}

\newcommand{\del}{\partial}
\def\ld{\Delta_d}

\def\S{\mathcal{S}_{_{0}}}
\def\Uz{\ \mathcal{U}_{_{0}}}
\def\Vz{\ \mathcal{V}_{_{0}}}
\def\Wz{\ \mathcal{W}_{_{0}}}

\def\L{\mathcal{L}}
\def\hb{{h_{_0}}}
\def\hbt{{\hat{h}_{_0}}}

\def\lgb{\varepsilon_{\pm}}

\def\ca{\scr{C}_{(\text{A})}}
\def\cb{\scr{C}_{(\text{B})}}

\def\ltwo{\ell_{\textsc{\ssmall 2}}^{\ \! \! 2}}
\def\lthree{\ell_{\textsc{\ssmall 3}}^{\ \! \! 4}}
\def\lfour{\ell_{\textsc{\ssmall 4}}^{\ \! \! 6}}

\newcommand{\beq}{\begin{equation}}
\newcommand{\eeq}{\end{equation}}
\newcommand{\beqs}{\begin{equation*}}
\newcommand{\eeqs}{\end{equation*}}

\newcommand{\bea}{\begin{eqnarray}}
\newcommand{\eea}{\end{eqnarray}}
\newcommand{\beas}{\begin{eqnarray*}}
\newcommand{\eeas}{\end{eqnarray*}}

\begin{document}

\begin{flushright}
\end{flushright}
\begin{center}

\vspace{1cm}
{\LARGE{\bf  On Swift Gravitons}}

\vspace{1cm}

\thispagestyle{empty}

{\large{Karim Benakli$^{1,2,a}$,
\let\thefootnote\relax\footnote{$^a$kbenakli@lpthe.jussieu.fr}
Shira Chapman$^{3,b}$ \footnote{$^b$schapman@perimeterinstitute.ca},
Luc Darm\' e$^{1,2,c}$ \footnote{$^c$darme@lpthe.jussieu.fr} and Yaron
Oz$^{4,d}$ \footnote{$^d$yaronoz@post.tau.ac.il}}}
\vspace{0.7cm}

{\emph{$^1$ Sorbonne Universit\'es, UPMC Univ Paris 06,UMR 7589, LPTHE,\\
F-75005, Paris, France \\
$^2$ CNRS, UMR 7589, LPTHE, F-75005, Paris, France \\
$^3$ Perimeter Institute for Theoretical Physics, Waterloo, Ontario N2L 2Y5, Canada\\
$^4$ Raymond and Beverly Sackler
School of Physics and Astronomy \\
Tel-Aviv University, Ramat-Aviv 69978, Israel}}

\end{center}
\vspace{0.7cm}

\abstract{We use the method of characteristics to study superluminal graviton (thereof called swift graviton) propagation in theories of higher curvature gravity of the form  $($Riemann$)^2$, $($Riemann$)^3$, $\nabla^2 ($Riemann$)^2$ and $($Riemann$)^4$. We consider a  \emph{pp}-wave background. When probed by gravitons with an appropriate polarisation, several of the gravitational theories under consideration exhibit characteristic hypersurfaces outside the flat spacetime light-cone. }

\newpage
\clearpage
\setcounter{page}{1}

\setcounter{tocdepth}{2}
\tableofcontents

\setcounter{footnote}{0}

\vspace{0.3cm}
\hspace{\fill}\rule{0.6\linewidth}{.2pt}\hspace{\fill}

\section{Introduction}

We consider extended gravity theories described by Lagrangians in $D$ spacetime dimensions of the form ($c=\hbar=1$):
\begin{equation}
\label{genericExtGrav}
 \scr{L} = \frac{1}{16 \pi \ell_P^{D-2}}\sqrt{-g} \big (  R + \ltwo \left[R^2\right] + \lthree \left[R^3\right]+ \lfour \left[R^4\right]\big ) ,
\end{equation}
where $g$ is the determinant of the metric $g_{\mu \nu}$, $\ell_P$ is the Planck length, $l_i$ are dimensionful coupling constants and $ \left[R^n\right]$ is a linear combination of contractions of $n$ Riemann tensors.\footnote{The $\left[R^3\right]$ term includes in addition to all possible contractions of three Riemann tensors also those of two covariant derivatives and two Riemann tensors.}
We will focus on theories with $\ell_i \gg \ell_P$, and consider the propagation of gravitons of energy $E\sim \ell_i^{-1}$, in which case  all the different terms in \eqref{genericExtGrav} are important.

Examples of \eqref{genericExtGrav} can be obtained from truncations of low energy effective actions of very weakly coupled string theory. In that case, the coefficients of the expansion are related to the string scale $\ell_s$,   $\ell_i \sim \ell_s$, and the hierarchy $\frac{\ell_P}{\ell_s} \ll 1$ is a consequence of the small string coupling. Such a truncation was shown in \cite{camanho_causality_2014} to lead to non-causality, in contrast to the case of the full fledged string theory. More precisely, for the case where the Lagrangian contains contractions of two or three Riemann tensors, the time-shift induced for  gravitons scattering on a ``shock-wave'' background \cite{dray_gravitational_1985} with an impact parameter $b\sim \ell_i$ has been computed. It was found that for a proper choice of the graviton polarisation the time-shift (the Shapiro time-delay in general relativity) can be turned into a time-advance effect. In the same paper, the time-advance effect was also obtained by considering a succession of graviton scattering events at impact parameters $b\sim \ell_i$.

How such time-advance effects arise has been further illustrated in different setups. The authors of~\cite{dappollonio_regge_2015} studied high-energy string-brane collisions. The authors found that the Shapiro time-delay can become negative when keeping only the contributions of massless states and suitably choosing the polarisations of the graviton in the initial state. This behaviour disappears when the full string theory amplitude is considered. The possibility of having a time advance  was also considered in \cite{papallo_graviton_2015} for  propagation of gravitons in the background of a small black hole solution of Einstein-Gauss-Bonnet theory.

In this work, we provide another illustration of the appearance of time-advances. We use the method of characteristics as in\cite{papallo_graviton_2015,Deser:2014hga,Deser:2014fta} to partially determine the causal cone for propagation of gravitons on a simple background solution of theories of the type \eqref{genericExtGrav}. We use a  \emph{pp}-wave background metric  solution induced by two identical parallel beams propagating at the speed of light in the same direction. The background metric can be taken to be a small perturbation of flat spacetime and we require that it is  a full solution of the extended gravity theories under consideration. We consider a probe graviton propagating right in the middle between the two beams such that it is not deflected. This means that we do not have access to the complete set of characteristic hypersurfaces but only to its restriction to signals propagating parallel to the beams. Because of this we cannot fully investigate the stability of our background solution in the various cases nor the hyperbolicity of our equations around this solution (for a more detailed discussion see e.g. \cite{Sarbach:2012pr,courant_methods_1962,Hormander:1983}).

Defining a time-shift on a background which is not Minkowski spacetime is a non-trivial task. Different theorems constraining the existence of ``fast travel'' by assuming various energy conditions can be found in the literature \cite{Gao:2000ga,Visser:1998ua,Palmer:2002vn,Olum:1998mu}. Each of them carries a different definition of ``fast travel'', which we refer to as ``swift'' propagation.\footnote{The word ``superluminal'' is usually associated with the presence of a causality violation which we do not prove here. Therefore, we introduce the name ``swift'' to avoid confusion.} For example~\cite{Gao:2000ga} focuses on the asymptotic structure of a given spacetime and shows, loosely speaking, that signals cannot travel faster than the speed allowed by the asymptotic causal structure. In~\cite{Visser:1998ua}, stationary perturbations of Minkowski spacetime in the harmonic gauge were considered. It was shown that  null geodesics of the total metric always lie inside the light-cone of flat spacetime. A similar and non-perturbative result was obtained in~\cite{Palmer:2002vn} by considering static spherically symmetric spacetimes. In our case, it is difficult to use the notion developed in~\cite{Gao:2000ga} as we do not have access to the complete causal structure of our theory. However, since our background is a small, stationary perturbation of flat spacetime, we use~\cite{Visser:1998ua} to define swiftness as the property that the characteristic hypersurfaces lie outside the flat spacetime light-cone.

We classify all possible extended gravity actions of the form $ \left[R^2\right]$, $ \left[R^3\right]$ for which our background is a solution. We show that in our setup $f(R)$ and Lovelock theories,  except for Gauss-Bonnet gravity, have the same causal structure as general relativity with no swift propagation.  However, we find that Gauss-Bonnet gravity as well as several other theories have characteristic hypersurfaces lying outside of the flat spacetime light-cone. Swiftness is therefore a common feature in extended gravity theories.
In $D>4$, we find that the sign of the coupling constants of the extended gravity theories are irrelevant in determining the presence of swift propagation on the background considered here. In contrast, for $D=4$, we observe that the swift nature of propagation depends on the signs of the coupling constants of the extended gravity theory considered.

Many of the theories we consider lead to equations of motion with more than two  derivatives acting on the probe graviton  and hence may be subject to Ostrogradsky instabilities. Common solutions include looking for theories with degenerate higher derivative terms like $f(R)$ theories (see~\cite{Woodard:2006nt}), restricting to actions with equations of motion with no more than two derivatives, like Lovelock theories, or adding constraints to remove the instabilities~\cite{Chen:2012au}. The swift propagation issue we discuss here is a priori unrelated and can further constrain the well-behaved extended gravity actions. For example, Gauss-Bonnet gravity, has equations of motion which are second order in derivatives, but does nonetheless exhibit swift propagation.

Our background is automatically a solution of $ \left[R^4\right]$ extended gravity theories. These theories often lead to equations of motion for the probe graviton which are fourth order in derivatives. While this is not a priori incompatible with causality, we show that such theories often lead to a degenerate causal cone and argue that this situation does not lead to a well-defined initial value problem. We present a simple way of curing this pathological case in $D>4$ theories by adding carefully chosen $[R^3]$ terms.

We work in a perturbative expansion in interactions of gravitons with a classical background.
When we restrict our equations of motion to be at second order in derivatives of the probe graviton we recover the structures observed by \cite{camanho_causality_2014} and \cite{dappollonio_regge_2015}.

This paper is organised as follow. In section \ref{sec:bkd}  we present our background which is a \emph{pp}-wave metric solution of general relativity. We also explain the choices that we make for the polarisations of the probe graviton and its propagation.
In section \ref{sec:charact} we detail our definition of  swift propagation. We give a brief review of the method of characteristics  and demonstrate it for the simple example of  $F^2 + RF^2$ that shows a swift behaviour as was already pointed out long ago in \cite{drummond_qed_1980} (for a recent discussion, see \cite{Hollowood:2015elj}).
In section \ref{sec:EOM}, we derive the equations of motion for the various theories under consideration. In section \ref{sec:suplm}, we study the implications of these equations to swift propagation. The analysis is restricted to the higher-dimensional cases $D>4$. The case of $D=4$ is discussed in section~\ref{sec:4D}.
In appendix~\ref{app:List} and appendix~\ref{app:rules}, we provide the detailed calculations behind our results.
In appendix~\ref{app:charact} and appendix~\ref{app:causality} we give more details about the method of characteristics and the notion of hyperbolicity.

\section{The Background Metric and Graviton Equations of Motion}

\label{sec:bkd}

In this section we construct a solution to Einstein equations describing the background metric induced by an axisymmetric source beam. We then consider a small perturbation around this background representing a propagating graviton.

\subsection{The Beam Metric}

We consider the following metric, axisymmetric around the $x_{D-1}$ axis:
\begin{equation}
\label{gbkd}
 ds^2 = -dudv + \hb(r) du^2 + \sum_{i=1}^{D-2} (dx_i)^2  .
\end{equation}
We use light-cone coordinates defined as $u = t-x_{D-1}$, $v = t+x_{D-1}$  where $t$ and $x_i$ denote the time and space coordinates, respectively. We also define the transverse distance: $r \equiv ({\sum_{i=1}^{D-2} (x_i)^2})^{\frac{1}{2}}$.

The only non zero Christoffel symbols are given by:
 \begin{align}
 \label{Chr_bkd}
  \Gamma^i_{~uu} = -\frac{1}{2} \d^i \hb, \qquad\Gamma^v_{~iu} = - \d^i \hb   ,
 \end{align}
 the non-zero components of the Riemann and Ricci tensors are:
\begin{equation}
\label{Riemannbkd}
 R_{iuju} = -\frac{1}{2}  \d_i \d_j \hb, \qquad
 R_{uu} = -\frac{1}{2} \d_i \d^i \hb   ,
\end{equation}
and the Ricci scalar vanishes.

The source is taken to be a ``Bonnor beam'' \cite{bonnor1969gravitational}, namely an infinitely long straight beam of radius $r_0$, propagating at the speed of light in the direction of the $v$ axis. It corresponds to the energy momentum tensor:
\begin{align}
\label{Tmunu}
T_{uu} =
\begin{cases}
\rho \qquad  r \leq r_0 \\
0  \qquad
 r > r_0
\end{cases},
\qquad
T^{vv} =
\begin{cases}
4\rho \qquad  r \leq r_0 \\
0  \qquad
 r > r_0
\end{cases} \ ,
\end{align}
while all the other components vanish and $\rho > 0$. 
The Einstein equations of motion reduce to:
\begin{align}
\label{EoMbkd}
  \d_i \d^i  \hb = - 16 \pi \ell_P^{D-2} \,  T_{u u } \ .
\end{align}
The solution to Eq.~(\ref{EoMbkd}) for $D>4$ is given  by:
\begin{align}
\label{hbkd}
 \hb(r) =  \begin{cases}
\displaystyle ~\frac{16 \pi \rho  \, \ell_P^{D-2}  \, r_0^2}{(D-4)(D-2)}   \left[\frac{r_0}{r}\right]^{D-4} & \qquad r>r_0
 \\[1.2em]
\displaystyle ~\frac{8 \pi \rho  \,  \ell_P^{D-2} \,   r_0^2}{(D-4)} \left[1-  \left(\frac{D-4}{D-2}\right)\frac{r^2}{r_0^2}  \right] & \qquad r\leq r_0
\end{cases}
  \ .
\end{align}
This solution belongs to a class of solutions called  ``plane fronted waves with parallel rays'' (or \emph{pp}-waves) and enjoys a superposition property, meaning that the linear sum of two parallel \emph{pp}-waves propagating in the same direction is still a solution. Away from the beam, at $r > r_0$, $\hb$ satisfies
\begin{equation}
\del_i \del^i \hb = 0 \ .
\end{equation}
It will be convenient to define $\Rb$ as
\begin{equation}\label{eq:DefRb}
  \Rb^{D-4}(\rho, r_0) \equiv \frac{16 \pi    }{(D-4)(D-2)} ~  \rho \,  \,  \ell_P^{D-2} \,   \,   r_0^{D-2} \ ,
\end{equation}
such that for $r>r_0$:
\begin{equation}\label{eq:theHb}
\hb = \displaystyle \left(\frac{\Rb}{r}\right)^{D-4} .
\end{equation}
In the limit  $r_0 \rightarrow 0$, the linear density of energy $\lambda_{\rho}$  is kept fixed by taking $\rho r_0^{D-2} \rightarrow \lambda_{\rho}  \frac{\Gamma(D/2)}{\pi^{(D-2)/2}}$. The background~\eqref{hbkd} then becomes:
\begin{equation}
\begin{split}
 \hb(x_i)  & \rightarrow ~  \frac{16 \pi \ell_P^{D-2} }{(D-4)(D-2)}  \  \left(\frac{\lambda_{\rho} \Gamma(\frac{D}{2})}{\pi^{\frac{D-2}{2}}}\right) \ \frac{1}{r^{D-4}} \\
& =~ \frac{4}{\pi^{(D-4)/2}} \Gamma \left(\frac{D-4}{2}\right) \frac{\lambda_{\rho} \ell_P^{D-2} }{r^{D-4}} \ ,
\end{split}
\end{equation}
which corresponds to:
\begin{align}
 T_{u u } =  \lambda_{\rho} \delta^{D-2}(x^i) \ .
\end{align}
Note that the shock-wave background of \cite{camanho_causality_2014} can be recovered by taking $\lambda_{\rho} = - P_u \delta(u)$  with $P_u$ the momentum of a single particle. Actually, our use of the metric \eqref{hbkd} was inspired by the fact that, in order to show a sizeable time advance, \cite{camanho_causality_2014} considered a set of successive scattering events. Such a stream can be approximated by  a source term with smoothed linear energy density of the type described above.

We choose the energy density and beam radius such that:
\begin{align}\label{eq:sizebkd}
 \hb \sim \ell_P^{D-2} \rho r_0^2 \ll 1 \ ,
\end{align}
so that we are considering a small perturbation to flat spacetime. We stress that our background is nonetheless in general a full solution of the extended gravity theories we will be considering, so we will not rely on \eqref{eq:sizebkd} to make a perturbative analysis.

\subsection{Equations of Motion for the Probe Graviton} \label{sec:pert}

We allow for a small perturbation around the background metric \eqref{gbkd} of the form:
\begin{equation}\label{eq:the_probe_graviton}
\delta g_{ij} = h_{ij} (u,v),
\end{equation}
describing a propagating ``probe'' graviton. We work in the light-cone gauge, $g_{vv} = g_{vi} = g_{ui} = 0$ (see e.g. \cite{Goroff:1983hc}). We will restrict the discussion to gravitons $h_{ij}(u,v)$ having transverse and traceless polarisation $\epsilon_{i j}$.\footnote{In $D=4$, such a restriction implies that the polarisations space is two-dimensional. However, the additional constraint~\eqref{constep} will further constrain the polarisation space to be one-dimensional. We discuss this issue in detail in Section~\ref{sec:4D}.} The tracelessness restriction simplifies the equations of motion but, in contrast to the case of flat spacetime, it is not enough to solve all the constraints given by the Einstein equations.

Appendix \ref{app:List} lists the metric components, the non vanishing Christoffel symbols, Riemann and Ricci tensor components and Ricci scalar up to first order in  the probe graviton contribution. Particularly useful for our calculations are:
\begin{equation}
\begin{split}
\label{Rbkd}
 & R_{iuju} = -\frac{1}{2} \hb_{,ij}  -\frac{1}{2} h_{ij, uu}  , \qquad  R_{ivjv} =  -\frac{1}{2} h_{ij, vv} ,
 \qquad  R_{iujv} =  -\frac{1}{2} h_{ij, uv}  ,
 \\
 & R_{ij} = 2  (h_{ij, uv} +  \hb h_{ij, vv}) ,
\quad
 R_{iu} =   \del^j\hb h_{ij, v}   ,
\quad
 R_{uu} =  -\frac{1}{2}(\hb_{,ii} - h^{ij} \hb_{,ij} ),\\
&~~~~~~~~~~~~~~~~~~~~~~~~~~~~~~~~~
 \qquad R  = 0  \ .
\end{split}
\end{equation}

For general relativity on this background, the Einstein equations $E_{\mu\nu}=0$ (away from the beam) take the form:
\begin{align}
& E_{ij} =  2 \left( h_{ij, uv} +  \hb h_{ij, vv}  \right) = 0, \label{eq:E_I_J}
\\
& E_{iu} = \del^j \hb h_{ij, v} = 0
, \label{eq:E_I_u}
\\
& E_{uu} =  -\frac{1}{2}(\hb_{,ii} - h^{ij} \hb_{,ij} ) = 0, \label{eq:E_u_u}
\end{align}
and the other components vanish identically.

Eq.~\eqref{eq:E_I_J} is the wave equation describing the propagation of the probe graviton and will be used to extract the causal cone of the theory using the method of characteristics. This will be discussed extensively in the next sections.

The gravitational  interaction of the graviton with the beam would deflect it, preventing $h_{ij}$ from being independent of the transverse coordinates.\footnote{In \cite{papallo_graviton_2015} where the scattering of gravitons off a black hole is studied, a zero-deflection trajectory can be found. The authors interpret it as the interplay of an attractive force from the usual Einstein gravity and a new -- repulsive -- force from the Gauss-Bonnet interaction. Such a zero-deflection trajectory in the presence of a single background source does not appear under our assumption. It would be interesting to see if we can observe it with more generic perturbations.} We introduce a second identical beam so that the graviton is propagating right in the middle between the two beams without deflection, making Eq. \eqref{eq:E_I_u} automatically satisfied.

Since any linear combination of \textit{pp}-waves sharing the same killing vector is also a solution of Einstein equations, we do not have to impose any condition on the distance between the beams to ensure they do not influence each other. The background line element is then given by:
\begin{equation}
 ds^2 = -dudv + (h^{(1)}_{_0} + h^{(2)}_{_0}) du^2 + \sum_{i=1}^{D-2} (dx_i)^2 \ ,
\end{equation}
where $h^{(1)}_{_0}$ and $h^{(2)}_{_0}$ are of the form (\ref{hbkd}) with the appropriate distances $r$ and $\hat{r}$. We concentrate on the behaviour of gravitons propagating between the two beams, such that the impact parameter $\vec{b}_1$ with respect to the first beam and $\vec{b}_2$ with respect  to the second one are opposite, $\vec{b}_1 + \vec{b}_2 = 0$. Particles moving in the $x_{D-1}$ direction in the middle between the two beams  have a non-deflected trajectory which is independent of the transverse directions. At these points of spacetime:
\begin{align}
\label{assump1}
 \left. \d_i (h^{(1)}_{_0} + h^{(2)}_{_0}) \right|_{\vec{b}_1 + \vec{b}_2= 0} = \left. \d_i \d_j \d_k(h^{(1)}_{_0} + h^{(2)}_{_0}) \right|_{\vec{b}_1 + \vec{b}_2= 0} = 0 \ .
\end{align}
As a consequence, any instance of a zero order Christoffel symbol which is not differentiated by a partial derivative vanishes. In what follow, we will always put ourselves in this configuration, and so, $\hb$ should be understood as $\hb\equiv h^{(1)}_{_0} + h^{(2)}_{_0}$.

Finally, Eq. \eqref{eq:E_u_u} is  a modification of the  zeroth order one Eq. (\ref{EoMbkd}) that determines our background. There, the additional first order term in $h^{ij}$ indicates that the probe graviton can back-react on the background metric unless:
\begin{align}
\label{assump2}
  h^{ij} \hb_{,ij} = (4-D)\frac{R_b^{D-4}}{b^D}(b^2 h^{ij} \delta_{ij} - (D-2) h^{ij} b_i b_j) = 0  \ .
\end{align}
Since $h^{ij}$ is traceless, this reduces to:
\begin{align}
\label{constep}
 \ep^{ij} b_i b_j = 0 \ .
\end{align}
Therefore, our equations describe the motion of a graviton propagating right in the middle between two beams. For this particular trajectory, one can find $D(D-3)/2 -1 $ traceless transverse polarisations which satisfy all the components of the Einstein equations, identically  except for $E_{ij}$ which describes the propagation along the trajectory.

In order to observe swift behaviour, we will see that we must require ``close fly-by'' of our test graviton, namely $b \sim \ell_n$. The condition that we look at gravitons propagating outside of our beam ($b > r_0$) then implies:
\begin{align}
 r_0 \lesssim  \ell_n \ ,
\end{align}
where $l_n$ are the coupling constants of equation \eqref{genericExtGrav}.

While in this section we have dealt with the Einstein-Hilbert action, we will show in section~\ref{sec:EOM} that the constraints \eqref{assump1} and \eqref{constep}  will remain sufficient to solve the $E_{uu}$ and $E_{ui}$ components of the equations of motion for the extended gravity theories under consideration.

\section{Characteristics and Causal Structure}
\label{sec:charact}

In this section, we define our notion of swift propagation and  explain the use of characteristics hypersurfaces to study the maximal propagation speed. To illustrate the procedure, we work out explicitly the examples of general relativity and of an $RF^2$ theory in subsection \ref{subsec:exm}.

\subsection{Time Shift and Swift Propagation}
\label{subsec:timeshift}

A criterion for superluminality, put forward in~\cite{Gao:2000ga} is that signals cannot travel faster than the speed allowed by the asymptotic causal structure of the background metric. Given a compact spacetime region $K$, this means that we can always take the start and end points for our signal far enough from $K$ such that the fastest path between the two does not cross $K$.
However this theorem requires the complete knowledge of the causal structure of the theory to find the fastest path. We do not have this information in the case of our background.
In  \cite{Visser:1998ua}, small perturbations of Minkowski spacetime in the harmonic gauge were considered. The authors show that null geodesics of the total metric, and therefore the paths of fastest signals, always lie inside the light-cone of flat spacetime. The background~\eqref{hbkd} satisfies the harmonic gauge condition so that the previous requirement should hold if $\hb \ll 1 $ at least for general relativity.

We chose our background such that $\hb$ goes to zero at infinity, which is obtained by taking the beams to be generated by a small density of energy as explained in the previous section, with a big but finite length so that our metric is locally a very good approximation.
In order to define ``superluminal''  or ``infraluminal'' speeds, we compare the propagation speed in a given background to the propagation speed in a flat Minkowski spacetime. This is equivalent to comparing the characteristic hypersurfaces with those asymptotically far away, in the limit $r\rightarrow \infty$, which are those of Minkowski spacetime.  Gravitons propagating faster than the Minkowski speed of light are then called ``swift gravitons''. See for instance a discussion in~\cite{Babichev:2007dw} for the relation between superluminality and violation of causality.

In $D>4$, we can proceed straightforwardly and define swiftness by comparing the characteristic hypersurfaces with the flat spacetime light-cone. Suppose that we find that one of the characteristic is of the form $(\xi_u = - \dc ,  \xi_v = 1) $ with $ | \dc | \ll 1 $, then swift propagation occurs for $\dc < 0$.

In $D=4$, the situation is more delicate
 and we need to use the fact that the beams have a finite length $L$. In the relevant limit where this length is very large compared to the other length scales in the problem, our background is of the form $ \hb \sim \log \frac{L}{r}$ where the beam length plays the role of an IR cut off. Because of the logarithmic growth of this metric, we remain in the limit $\hb \ll 1 $ for all practical purposes. We find that in that case,  $\dc$ is of the form $\dc = a (  \log \frac{\Lambda}{r} + \frac{N \ell^m}{r^m})$ where $N$ is a $\scr{O} (1)$ number which can be either positive or negative depending on the polarisation of the probe graviton, $a$ is a constant depending on the source beam, $\ell^m$ is the coupling constant of the extended gravity terms and $m \geqslant 4$ an integer which depends on the extended gravity theory. As was already pointed out in~\cite{camanho_causality_2014}, it is always possible to find a setup in which the logarithm is overtaken by the other terms such that we can focus on the sign of their contribution to determine if we have swift propagation.

\subsection{Method of Characteristics}\label{subsec:methCar}

In order to study the causal structure of the theories of extended gravity under consideration, we will study their characteristic hypersurfaces. Those can be deduced from the equations of motion for a probe graviton in the background \eqref{gbkd}.
 In the following subsection we explain how this is done.

Consider a linear equation of motion for a scalar field $\phi$ of the form
\begin{equation}
 P(\d) \phi = 0 \ ,
\end{equation}
where $P$ is a polynomial in $\d$ of order $n$ with constant coefficients. We denote by $P_n$ the truncation of this polynomial to its highest derivative terms in $P$ ($P_n$ is often called the principal part of $P$). Each 1-form $\xi$ such that:
\begin{equation}
\label{eq:char}
P_n( \xi ) = 0 \ ,
\end{equation}
is normal to a codimension-1 hypersurface called a characteristic hypersurface.
On characteristic hypersurfaces a full set of initial Cauchy data including all inner pointing derivatives and the first $n-1$ outward pointing derivatives does not fully determine the
value of the $n$-th order outward pointing derivative. In particular, the $n$-th order outward pointing derivative can be discontinuous, allowing for the propagation of shock waves. For illustrative purposes, we explain the idea behind the method of characteristic for a scalar field in appendix \ref{app:charact}.

Since characteristic hypersurfaces are boundaries of the causal development of an initial hypersurface $\scr{I}$, they give the edges of the region where the physics is fixed by the initial data on $\scr{I}$. As a consequence of these properties, the characteristic hypersurfaces also determine the trajectories of signals with the highest propagation speed in the theory considered.

To use the method of characteristics with the equations of motion for our probe graviton, one should take into account the following two subtleties.
First, the coefficients of the PDEs could in principle be (non-constant) functions of the spacetime coordinates. However, since we are considering a setup in which the test graviton has a constant impact parameter, it remains static in the transverse directions and therefore its equations of motion admit constant coefficients.
Second, we are considering a tensor field, hence (\ref{eq:char}) should be modified to a determinant of the set of characteristic equations corresponding to the different polarisations of the probe graviton. However, we will take the
basis of polarisation such that the system of equations of motion becomes diagonal.\footnote{This means that the Einstein equations do not mix different polarisations and we have the form: $E_{ij} \propto \epsilon_{ij}$.} We can therefore consider each polarisation individually and look for those exhibiting a swift behaviour. The detailed choices of polarisations that we take are described at the beginning of section \ref{sec:suplm}.

Since we are only considering the equations of motion in the $(u,v)$ plane, we cannot check that the full equations of motion are hyperbolic.\footnote{The precise definition of hyperbolicity is given in appendix \ref{app:causality}. In\cite{reall_causality_2014}, it was shown that Lovelock theories, for instance, are hyperbolic around pp-wave solutions.} This property is required to make sure that the notion of causal structure is well-defined, as we review in Appendix~\ref{app:causality}. If we assume that the causal cone does exist, then our results give locally its projection to the $(u,v)$ plane. We therefore only investigate swift  propagation of gravitons restricted to this plane.

We find that the projected causal cone  can be degenerate in theories with terms of the form $\text{(Riemann)}^4$ where the characteristic equation has one root with degeneracy four. In this situation, we use a theorem stated in appendix \ref{app:causality} to prove that the full equations of motion cannot be hyperbolic with respect to the time direction when expanded around our background, and therefore the initial value problem is not well-posed. The theorem relates the degeneracy of the roots of different truncations of the characteristic polynomial.
 We leave the complete proof to the appendix. Three comments are in order. First, in this degenerate case, the hyperbolicity  depends on the complete equations, and not only on the terms at highest order in derivatives. Second, in our proof, we rely on the fact that our action always includes the Einstein-Hilbert term. Third, the non-hyperbolicity is a local statement because we are only considering probe gravitons propagating in the middle between the two beams.

\subsection{Characteristic Equations: Structure and Examples}
\label{subsec:exm}
We are interested in equations of motion of the form:
\begin{equation}
\label{eq:generalstructure}
 \d_u \d_v h + a \d_v^2 h+ b \d_v^4 h + c \d_u \d_v^3 h  = 0 \ ,
\end{equation}
with three different cases:
\begin{enumerate}
\item When $c=b=0$, we have a situation similar to Einstein gravity. Our equations of motion are second order in derivatives of the probe graviton and we have two distinct roots for the characteristic equation. Swift propagation occurs if $a<0$.
\item When $b \neq 0$ and $ c = 0$, we have only one characteristic $\xi_v = 0$. The causal cone is degenerate, we are in the situation discussed in the previous subsection.
\item When $b \neq 0$ and $ c \neq 0$, we have a causal cone whose boundaries are the two characteristics given by $\xi_u =  -\xi_v b/c$ and $\xi_v=0$. Swift propagation occurs if $b/c <0$.
\end{enumerate}

We start by studying the example of a graviton propagating in Einstein gravity described by the equations of motion \eqref{eq:E_I_J}. The characteristic equations read:
\begin{align}
  2\xi_u  \xi_v   \ep^{ij} + 2\hb \xi_v^2 \ep^{ij} = 0 \ .
\end{align}
These equations are diagonal in the polarisations (i.e. they satisfy $E_{ij} \propto \ep_{ij}$). Therefore, we can consider each polarisation separately. The equations have two solutions, independent of the polarisation of the graviton, giving two characteristic hypersurfaces:
\begin{align}
\label{eq:charEinstein}
 (\xi_u = 1, \xi_v = 0)  \text{   and   }  (\xi_u = - \frac{\hb}{1+\hb^2} ,  \xi_v = \frac{1}{1+\hb^2})  .
\end{align}
Since $\hb > 0$, gravitons propagating in the negative $x_{D-1}$ direction suffer a time delay and their propagation is infraluminal. This is the Shapiro time delay. Graviton propagating in the positive $x_{D-1}$ directions remains unaffected.

As a second example we consider an abelian gauge field in $D>4$ non-minimally coupled to gravity:
\begin{align}
 \scr{L} = \sqrt{-g} (\frac{1}{16 \pi \ell_P^{D-2}} R + \frac{1}{4}F_{\mu\nu}F^{\mu\nu} + \frac{1}{4} \ell_R^2 R^{\mu \nu \rho \s} F_{\mu \nu} F_{\rho \s}) \ .
\end{align}
We suppose that the background metric is of the form ~(\ref{gbkd}) and is induced by two parallel beams. We are interested in a particle propagating in the middle between the two beams. The equations of motion for $ A^\mu$ simplify to:
\begin{align}
 \d_\rho \d^\rho A^\mu -  \d^\mu \d_\rho A^\rho - 2 \ell_R^2 R^{\mu \nu \rho \s} \d_\nu \d_\rho A_\s = 0 \ .
\end{align}
We focus on a field $A^\mu$ with only transverse modes and suppose that it does not depend on the transverse coordinates. The only relevant components of the equation of motion are:
\begin{align}
 \d_u \d_v A^i +  (\delta^{i j} \hb + \ell_R^2 \d^i \d^j \hb) \d_v^2 A_j   = 0 \ .
\end{align}
The unit vector $\xi^\mu$ normal to a characteristic hypersurface satisfies:
\begin{align}
\xi_u \xi_v A^i +  (\delta^{ij} \hb + \ell_R^2 \d^i \d^j \hb ) \xi_v^2  A_j   = 0 \ .
\end{align}
We choose the impact parameter $\vec{b} = (b, 0 , \dots)$ and we consider two choices of polarisations:
\begin{itemize}[label=--]
\item (A): $A^{1}$ non zero and all the other components vanish,
\item (B): $A^{2}$ non zero and all the other components vanish.
\end{itemize}
We get:
\begin{align}
  \xi_v ( \xi_u + \scr{C} \xi_v )  = 0  \ ,
\end{align}
where $\scr{C}\in\{\ca,\cb \}$ depends on the polarisation of $A^i$:
\begin{align}
\ca =   \left(\frac{\Rb}{b }\right)^{D-4} \left[ 1 + \ell_R^2\frac{(D-4)(D-3)}{b^2} \right] \ ,
\end{align}
and
\begin{align}
\cb =   \left(\frac{\Rb}{b }\right)^{D-4} \left[ 1 - \ell_R^2\frac{D-4}{b^2} \right] \ .
\end{align}
Taking $b$ small enough such that $\displaystyle \frac{\ell_R}{b} > 1$, $\ca$ and $\cb$ have opposite signs. Therefore we can have $\scr{C} < 0$
regardless of the sign of $\ell_R^2$.\footnote{Note, that the notation $\ell_R^2$ is made to match the conventions in the literature. We do not need to assume it is of positive sign for our argument in this subsection.} The two characteristic hypersurfaces are given by:
\begin{align}
 (\xi_u = 1, \xi_v = 0)  \text{   and   }  (\xi_u = - \frac{\scr{C}}{1+\scr{C}^2}  ,  \xi_v = \frac{1}{1+\scr{C}^2})  \ .
\end{align}
Hence, when $\scr{C}<0$, $A^i$ propagates toward decreasing $x_{D-1}$ at a superluminal speed.
Note that the relative size of $R_b$ compared to $b$, affects only  the strength of the time shift effect. Its advance or delay nature is determined by the polarisation of $A^i$ and by the impact parameter.  Therefore, swift propagation can happen even in the weak curvature regime when $\left(\frac{\Rb}{b }\right)^{D-4} \ll  1 $.

\section{Equations of Motion in Theories of Extended Gravity}\label{sec:EOM}

In this section, we study theories of extended gravity in $D>4$. For the cases in which the background \eqref{gbkd} with $\hb$ defined in \eqref{hbkd} remains a solution, we derive the equations of motion  for a propagating probe graviton of the form \eqref{eq:the_probe_graviton}.
We will fully classify the few cases in which this condition does not hold and will not  make statements about their causal structure.\footnote{These are the terms $R_{ab} R^{ab}$, $R_{abcd} R^{abcd}$ and $R_{ab} \nabla^2 R^{ab}$ (see appendix \ref{app:rules} Table \ref{table:EOM_0}). However, it is possible to show that gravitons propagating around flat background in these theories have equations of motion with terms which contain four or six  derivatives acting on the probe graviton, so that these theories are very likely to develop Ostrogradsky instabilities.}
 Some of the results of this section were derived using \textit{Mathematica} and the rules listed in appendix \ref{app:rules}.

\subsection{General Considerations}\label{subsec:Gen_con}

We start by general statements regarding those components of the equations of motion that involve indices which are not transverse.

In subsection \ref{sec:pert} we have seen that the $E_{uu}$ component of the equations of motion in Einstein gravity is automatically satisfied for traceless transverse polarisations when we require:
\begin{align}\label{pol_cons2}
  \ep^{ij} b_i b_j = 0 .
\end{align}
The same is true for $E_{uu}$, $E_{uv}$ and $E_{vv}$ in higher curvature gravity as they contain factors of the form:
\begin{align}
h_0^l |b|^{2m} b^i b^j \d_{u,v}^n h_{ij} \ ,
\end{align}
where $m$, $k$ and $l$ are natural numbers, and $\d_{u,v}^n$ stands for any number of $u$ and $v$ derivatives. These factors vanish when (\ref{pol_cons2}) is imposed.

The $E_{ui}$ and $E_{vi}$ equations of motion must involve at least one $\hb$ acted on by an  odd number of derivatives. This is since $h_{ij}$ indices come in even numbers. Therefore, the contributions of the two beams always cancel each other when the graviton propagates in the middle between them (see Eq. \eqref{assump1}) and the equations are identically satisfied.

We conclude that for our two-beams  \emph{pp}-wave background with the additional condition \eqref{pol_cons2}, all equations of motion are automatically satisfied except for $E_{ij}$ which we study next.

\subsection{Equations of Motion in Curvature Squared Gravity Theories}

There are three possible curvature squared terms which we can consider: $R^2$, $ R^{\mu \nu} R_{\mu \nu}$ and $R^{\mu \nu \rho \s}R_{\mu \nu \rho \s}$.
We list their contributions to the background equations of motion and those of the probe graviton in Table \ref{table:resR2}. Since we want to keep the form \eqref{hbkd} for $\hb$, we require that the zeroth order background equations remain the same as those of general relativity.  The actions satisfying this condition can be read from Table \ref{table:resR2} and include $f(R)$ gravity, Einstein-Gauss-Bonnet and any linear combination thereof.
 The equations of motion of all those theories are found to be either vanishing or proportional to:
 \begin{equation}\label{eq:defT0}
 \mathcal{T}_{{}_{0}}^{ij} \equiv
 \hb_,^{\phantom{,} k (i} h^{j)}_{\phantom{j)} k , vv} \ ,
 \end{equation}
at first order in the probe graviton contribution.


\begin{table}[!ht]
\begin{center}
\begin{tabular}{ | c | c | c |}
   \hline
  Action & Background equations & EoM at first order in $h^{ij}$ \\
   \hline
   \rule{0pt}{3ex}
$  R^2$ &   $ 0$ & $0$ \\[0.4em]
$  R_{ab}  R^{ab}$ &   $ -\frac{1}{2} \vec{\d}^4 \hb $ & $\textcolor{orange}{ -\frac{1}{2} \ld^2 h^{ij}} $ \\[0.4em]
$ R_{abcd}   R^{abcd} $  &  $ -2 \vec{\d}^4 \hb $ & $ \textcolor{orange}{-2  \ld^2 h^{ij}} + 8 \hb_{,}^{  k  ( j } ~ \d_v^2 h^{i )   }_{~ k }  $ \\[0.4em]
  \hline
 \end{tabular}
 \caption{Contributions of the possible $[R^2]$ actions to the zeroth and first order equations of motion. Orange color indicates fourth order derivatives acting on $h^{ij}$. We have defined  $\ld \equiv -4 ( \d_u \d_v + \hb \d_v ^2 )$.}
\label{table:resR2}
\end{center}
\end{table}

The Einstein-Gauss-Bonnet gravity is a combination whose equations of motion are at most second order in derivatives of the probe graviton. The relevant Lagrangian is given by
\begin{align}
 \scr{L} = \frac{1}{16 \pi \ell_P^{D-2}}\sqrt{-g} \big [R + \lgb \ltwo (R^2 - 4 R^{\mu \nu} R_{\mu \nu} +  R^{\mu \nu \rho \s}R_{\mu \nu \rho \s} ) \big ] \ ,
\end{align}
where $\lgb$ stands for the sign of the Gauss-Bonnet term, and  the corresponding equations of motion take the form:
\begin{align}
 G_{\mu \nu} + \lgb \ltwo H_{\mu \nu} = 0 \ ,
\end{align}
where:
\begin{align}
  G_{\mu \nu} = R_{\mu \nu} - \frac{1}{2}g_{\mu \nu} R \ ,
\end{align}
is the usual Einstein tensor and:
\begin{equation}
\begin{split}
  H_{\mu \nu} =~ &  2(R R_{\mu \nu} - 2R_{\mu}^{~ \alpha} R_{\alpha \nu} -  2 R^{\alpha \beta}R_{\mu \alpha \nu \beta} + R_{\mu}^{~\alpha \beta \gamma} R_{\nu \alpha \beta \gamma}) \\
 & - \frac{1}{2} g_{\mu \nu} (R^2 - 4 R^{\alpha \beta} R_{\alpha \beta} +  R^{\alpha \beta \gamma \delta}R_{\alpha \beta \gamma \delta})\ ,
\end{split}
\end{equation}
is the additional Gauss-Bonnet contribution.
One can read the transverse equations of motion for the probe graviton from Table \ref{table:resR2} and obtain:
\begin{align}\label{eq:EOM_GB}
  E_{ij} = h_{ij, uv} +  \hb  h_{ij, vv} + 4 \lgb \ltwo~  \hb_{,k (i} ~h^{~ k}_{j)~,vv} = 0 \ .
\end{align}
We will study the characteristics hypersurfaces for this equation in the next section.

\subsection{Equations of Motion in $R^3$, $R\nabla^2 R$ and $R^4$ Theories}

In order to obtain the equations of motion for actions containing contributions of the forms $R^3$, $R\nabla^2 R$ and $R^4$ we used xAct \cite{martin-garcia_invar_2007,martin-garcia_invar_2008,martin-garcia_xperm:_2008} and xTras \cite{nutma_xtras:_2014},
tensor algebra packages for \emph{Mathematica}. We followed the steps bellow:
\begin{enumerate}
\item Generate all possible contractions of three Riemann tensors, four Riemann tensors, and two Riemann tensor and two covariant derivatives.
\item Select an independent basis of actions, taking into account integration by parts and various geometric identities.
\item Compute the full equations of motion for each of the above actions using xTras.
\item Check if the background equations  defining $\hb$ are modified.
\item Expand around $\hb$ given by \eqref{hbkd} and use the rules detailed in Appendix \ref{app:rules} to simplify and evaluate these equations of motion at first order in perturbation theory.
\end{enumerate}
Most of the rules derived in Appendix \ref{app:rules} are due to the peculiar form of our background and the possible number of instances and placements of the $u$ and $v$ indices.

We list out results for the equations of motion of $R\nabla^2R$ and $R^3$ actions in Table \ref{table:resR3}.
We obtain that all contributions of dimension six to the equations of motion of the probe graviton are linear combinations of two possible terms:
\begin{equation}\label{eq:defT1}
\mathcal{T}_{{}_{1}}^{ij} \equiv \hb_{,}^{ ~ k (j } ~ \ld \d_v^2 h^{i)  }_{~ k } \ ,
\end{equation}
and
\begin{equation}\label{eq:defS0}
\mathcal{S}_{{}_{0}}^{ij} \equiv \hb_{,}^{ ~ i j k l } ~ \d_v^2 h_{ k l } \ ,
\end{equation}
where we have defined the derivative operator $\ld$ by:
\begin{equation}
\ld = -4 ( \d_u \d_v + \hb \d_v \d_v ) \ .
\end{equation}
$\mathcal{T}_{{}_{1}}^{ij}$ is simply the $\ld$ derivative of the $\mathcal{T}_{{}_{0}}^{ij}$ term that we already encountered in the context of the Gauss-Bonnet action. We will study both contributions in the next section and show that they lead to  swift propagation for certain polarisations of the graviton.\footnote{In Table~\ref{table:resR3}, we also have a $ \ld^3 h_{ i j  } $ contribution, however, the action term which generates it, $R^{ab}  \nabla_c \nabla^c R_{ab}$, also modifies the background equations of motion, rendering our analysis irrelevant.}
\begin{table}[!ht]
\begin{center}
\begin{tabular}{ | c | c | c |}
   \hline
  Action &Background equations & EoM at first order in $h^{ij}$ \\
   \hline
   \rule{0pt}{3ex}
$R_{a}^{~c}R^{ab}R_{bc} $ & $0$ & $0$  \\[0.4em]
$R R^{ab}R_{ab} $ & $0$ & $0$  \\[0.4em]
$R^3 $ & $0$ & $0$  \\[0.4em]
$R^{ab}  R^{cd} R_{acbd} $ &  $ 0 $ & $  0$\\[0.4em]
$R  R_{abcd}  R^{abcd}$ & $0$ &  $ - 8 \hb_{,}^{ ~ i j k l } ~ \d_v^2 h_{ k l } $ \\[0.4em]
$R^{a b}   R_{a}^{~cde}   R_{bcde}$   & $0$& $ -2 \hb_{,}^{ ~ i j k l } ~ \d_v^2 h_{ k l } \textcolor{orange}{~+~ 4  \hb_{,}^{ ~  k (i } ~ \Delta_d \d_v^2 h_{ k }^{\phantom{k} j )}}$ \\[0.4em]
$R_{a ~ c ~}^{~ e~ f}R^{a b c d}   R_{bedf}$   &$ 0$& $  3 \hb_{,}^{ ~ i j k l } ~ \d_v^2 h_{ k l }  $  \\[0.4em]
$ R_{ab}^{ ~~ef}   R^{abcd}   R_{cdef}$  & $0$ & $  \textcolor{orange}{24  \hb_{,}^{ ~  k (i } ~ \Delta_d \d_v^2 h_{ k }^{\phantom{k} j )}}$ \\[0.4em]
 $R  \nabla_c \nabla^c R $ &  $0$ & $ 0$\\[0.4em]
$R^{ab}  \nabla_c \nabla^c R_{ab} $ &  $-\frac{1}{2} \vec{\d}^6 \hb $ & $ \textcolor{darkred}{-\frac{1}{2} \ld^3 h^{ij}} $\\[0.4em]
  \hline
 \end{tabular}
 \caption{ Various contributions from the possible dimension-six terms constructed from Riemann tensors and covariant derivatives.  Orange color indicates the presence of fourth order derivatives acting on $h^{ij}$ and red the presence of sixth order derivatives.}
\label{table:resR3}
\end{center}
\end{table}

Similarly, we can consider actions built from four Riemann tensors. The background equations are not modified in this case. All actions including at least one Ricci tensor or Ricci scalar have equations of motion which are identically satisfied for the probe graviton. Our results for the equations of motion of the test graviton are summarised in Table \ref{table:R4}.

There are three possible contributions to the equations of motion of $[R^4]$ terms, each with four derivatives acting on $h^{ij}$ given by:
\begin{align}\label{eq:defU0}
\mathcal{U}_{{}_{0}}^{(ij)} \equiv   \hb_{,}^{~lk} \hb_{, k}^{~~ (i} h^{j )}_{~~ l , v^4} \ ,
\end{align}
\begin{align}\label{eq:defW0}
\mathcal{W}_{{}_{0}}^{ij} \equiv 	  \hb_{,}^{~ik} \hb_{,}^{~jl} h_{kl , v^4} \ ,
\end{align}
and
\begin{align}\label{eq:defV0}
\mathcal{V}_{{}_{0}}^{ij} \equiv \hb_{,}^{~kl} \hb_{,kl} h^{ij}_{~~ , v^4}\ .
\end{align}

\begin{table}[!ht]
\begin{center}
\begin{tabular}{ | c | c |}
   \hline
  Action  & EoM at first order in $h^{ij}$ \\
   \hline
   \rule{0pt}{3ex}
$R_{a b }^{~ ~ ef}R^{abcd} R_{ce}^{~~hg} R_{dfhg} $ & $  - 32 \hb_{,}^{~lk} \hb_{, k}^{~ (i} \d_v^4 h^{j )}_{~~ l }$  \\[0.4em]
$R_{a ~ c }^{~ e~ f}R^{abcd} R_{b~e}^{~h~g} R_{dhfg} $ & $  - 8 \hb_{,}^{~lk} \hb_{, k}^{~ (i} \d_v^4 h^{j )}_{~~ l } - 4 \hb_{,}^{~kl} \hb_{, kl} \d_v^4 h^{ij}_{~~ }$  \\[0.4em]
$R_{a b }^{~ ~ ef}R^{abcd} R_{c~e}^{~h~g} R_{dhfg} $ & $ 0 $   \\[0.4em]
$R_{a b }^{~ ~ ef}R^{abcd} R_{cd}^{~~hg} R_{efhg} $ & $- 64  \hb_{,}^{~ik} \hb_{,}^{~jl} \d_v^4 h_{kl } $   \\[0.4em]
$R_{a b c }^{~ ~ ~e}R^{abcd} R_{d}^{~fhg} R_{efhg} $ & $ - 16  \hb_{,}^{~ik} \hb_{,}^{~jl} \d_v^4 h_{kl } - 16 \hb_{,}^{~lk} \hb_{, k}^{~ (i} \d_v^4 h^{j )}_{~~ l }$  \\[0.4em]
$R_{a~ c }^{~ e~ f}R^{abcd} R_{b~d}^{~h~g} R_{ehfg} $ & $ - 8  \hb_{,}^{~ik} \hb_{,}^{~jl} \d_v^4 h_{kl } - 8  \hb_{,}^{~kl} \hb_{,kl} \d_v^4 h^{ij}_{~~ }$  \\[0.4em]
$R_{a b c d } R^{abcd} R_{efhg} R^{efhg} $ & $0$  \\[0.4em]
  \hline
 \end{tabular}
 \caption{ Various contributions from the possible $[R^4]$ terms. The contributions to the background equations of motion vanish.}
\label{table:R4}
\end{center}
\end{table}

As an example we consider the third and fourth order Lovelock theories.
 \begin{equation}
\label{eq:lov3}
\begin{split}
  &\L_{3} = \sqrt{-g}  \left(  R^3  + 16 R_{a}^{~c} R_{ab}R^{bc} + 24  R_{ab} R^{cd} R_{acbd}+3  R R^{abcd} R_{abcd}  \right. \\
 & ~~\left. - 12 R R_{ab}R^{ab} - 24  R^{ab} R_a^{~cde} R_{bcde} + 8  R_{a~c}^{~e~f} R^{abcd} R_{bfde} +   2  R_{ab}^{~~ef} R^{abcd} R_{cdef} \right),
 \end{split}
 \end{equation}
and
 \begin{equation}\label{eq:lov4}
 \begin{split}
  \L_4 = \sqrt{-g}&  \left(96 R_{a ~ c }^{~e~f}R^{abcd}R_{b ~ e }^{~h~i}R_{dhfi} + 96 R_{ab}^{~~ef} R^{abcd} R_{c ~ e }^{~h~i}R_{difh} \right. \\
 & - 6 R_{ab}^{~~ef} R^{abcd} R_{c d }^{~~hi}R_{efhi}+ 48 R_{abc}^{~~~e} R^{abcd} R_{d }^{~fhi}R_{efhi} \\
  & \left. -48 R_{a ~ c }^{~e~f}R^{abcd}R_{b ~ d }^{~h~i}R_{ehfi} - 3 \left(R_{abcd}R^{abcd}\right)^2 ~\right)  + \scr{F}(R^{ab}, R).
  \end{split}
 \end{equation}
Using the results of Tables \ref{table:resR3} and \ref{table:R4} we find that both the contributions to the background equations of motion and those to the equations of motion for the test graviton vanish. This is expected since the $E_{ij}$ components of the equations of motion in Lovelock theories take the form:
\begin{align}
 \sum_{n \ge 2} \lambda_n \delta^{i \rho_1 \ldots \rho_{2p}}_{j \s_1 \ldots \s_{2p}} R_{\rho_1 \rho_2}{}^{\s_1 \s_2} \ldots R_{\rho_{2p-1} \rho_{2p}}{}^{\s_{2p-1} \s_{2p}}
\end{align}
where $\delta$ is the generalised Kronecker delta. In particular, for $n>2$, we must have at least two Riemann tensors at zeroth order in perturbation theory and therefore must absorb four upper $u$ / lower $v$ indices. However, the generalised Kronecker delta can only absorb two upper $u$ / lower $v$ indices since it is fully antisymmetric, and so, the contributions from Lovelock terms of order $n>2$ vanish.

\section{Swift Behaviour}\label{sec:suplm}

In this section, we use the equations of motion derived in section \ref{sec:EOM} to find the characteristics hypersurfaces for the test graviton and check for swift propagation. We only consider cases with $D>4$. We start by explaining our choice of the basis of polarisations for the probe graviton.

\subsection{Choosing a Basis of Polarisations}

Without loss of generality, let us take the impact parameter to be $\vec{b} = (b, 0 , \dots)$ where  we label the transverse coordinates by $x_i, ~~i=1,2,3, \ldots$. The index structure forces all contributions to our equations of motion to be of the form:
\begin{equation}
\label{eq:genform}
F \epsilon^{ij} + G b^k b^{(i} \ep^{j)}_{\phantom{j} k} \ ,
\end{equation}
where $F$ and $G$ may contain $v$ and $u$  derivatives. Given the restriction \eqref{constep} and the fact that the polarisation is traceless, for gravitons propagating in the middle between the two beams, we can use the following basis of polarisations:
\begin{itemize} [label=--]
\item $\oplus$ polarisations -- these are polarisations of the form $\ep^{aa} = - \ep^{bb} = 1/\sqrt{2}$ with the requirement  that $a,b \neq 1$ (as a consequence of \eqref{constep}) and all other components vanish.
\item $\otimes$ polarisations -- these are polarisations  of the form $\ep^{ab} = \ep^{ba} = 1/\sqrt{2}$ and all other components vanish.
\end{itemize}
It is then easy to see that the second term of \eqref{eq:genform} vanishes for $\oplus$ polarisations and that it is non zero only for $\otimes$ polarisations of the form $\ep^{1b} = \ep^{b1} = 1/\sqrt{2}$.
Hence we identify two different classes of $\otimes$ polarisation:
those for which the second term of \eqref{eq:genform} vanishes and those for which it does not. Picking one representative for each class, we have:
\begin{itemize}[label=--]
\item Class (A):  $\otimes$ polarisation with  $\ep^{2 1} = \ep^{1 2 } = 1/\sqrt{2}$.
\item Class (B): $\otimes$ polarisation with  $\ep^{2 3} = \ep^{3 2 } = 1/\sqrt{2}$ .
\end{itemize}
Notice that the operator $F \delta^{ik} \delta^{jl}  + G b^k b^{(i}\delta^{j) l }$ is diagonal for this basis of polarisations (i.e. $E_{ij}\sim \epsilon_{ij}$ since $\epsilon^{k(i } b^{j)}b_k= b^2 \epsilon^{ij}/2$ for class (A) and $0$ for class (B)).
 We are therefore in the case discussed in section~\ref{sec:charact} and we can consider the characteristic equation for each type of polarisation separately.

\subsection{Swift Behaviour of $R^2$, $R^3$ and  $R\nabla^2 R$ Actions}\label{sec:classification}

Since the equations of motion are linear in the probe graviton it is possible to study the characteristic equations of each type of contribution to the equations of motion separately. We use $\ca$ and $\cb$ to denote contribution to the characteristic equations for gravitons with polarisations (A) and (B), respectively.

First, let us consider the contribution to the characteristic equations from the Einstein-Hilbert term $\mathcal{E}_{ij} = -\frac{1}{2}\ld h_{ij} $ of  Eq. \eqref{eq:E_I_J}:
\begin{equation}
\scr{C}_{\text{(A)/(B)}} \left[\mathcal{E}_{ij} \right]
=  2 \xi_v \left( \xi_u + \left(\frac{\Rb}{b }\right)^{D-4}\xi_v\right) \ep_{ij} \ .
\end{equation}
As we have already seen before, the characteristics hypersurfaces are given by Eq. \eqref{eq:charEinstein} and indicates  infraluminal propagation (Shapiro time delay) independent  of the choice polarisation.

We move next to terms which are quadratic in derivatives of the probe graviton. We have seen in the previous section that these are the $
\mathcal{T}_{{}_{0}}^{ij} =
\hb_,^{\phantom{,} k (i} h^{j)}_{\phantom{j)} k , vv}$ and  $\mathcal{S}_{{}_{0}}^{ij} = \hb_,^{\phantom{,} i j k l} h_{ k l , vv}$ terms of Eqs. \eqref{eq:defT0} and \eqref{eq:defS0}. Their  contributions to the characteristic equations are given by:
\begin{align}
\begin{split}
\ca \left[ \mathcal{T}_{{}_{0}}^{ij}  \right] & ~=~ \epsilon_{ij} \left(\frac{\Rb}{b }\right)^{D-4} \frac{ (D-4)^2}{2 b^2} \xi_v^2,\\
\cb \left[ \mathcal{T}_{{}_{0}}^{ij}   \right] & ~=~ -\epsilon_{ij} \left(\frac{\Rb}{b }\right)^{D-4} \frac{(D-4)}{ b^2} \xi_v^2 ,
\end{split}
\end{align}
and
\begin{align}
\begin{split}
\ca \left[ \mathcal{S}_{{}_{0}}^{ij} \right] & ~=~ - \epsilon_{ij} \left(\frac{\Rb}{b }\right)^{D-4}\frac{2 (D-4)(D-2)(D-1)}{ b^{4}} \xi_v^2 ,\\
\cb \left[\mathcal{S}_{{}_{0}}^{ij} \right] & ~=~ \epsilon_{ij} \left(\frac{\Rb}{b }\right)^{D-4} \frac{2  (D-4)(D-2)}{ b^{4}} \xi_v^2. \end{split}
\end{align}
Note that the signs of these contributions depend on the polarisation of the probe graviton.

We illustrate the swift behaviour of the probe graviton by studying two examples in details. These are Gauss-Bonnet gravity and an action of the form $R - \frac{\lambda_3 \ell_{\!\mathsmaller{3}}^4}{8}R  R_{abcd}  R^{abcd}$.
For Gauss-Bonnet gravity, we obtained the following equations of motion (see Eq. \eqref{eq:EOM_GB}):
\begin{align}
   h_{ij, uv} +  \hb  h_{ij, vv} + 4  \lgb \ell_{\! \mathsmaller{2}}^2 \hb_{,k (i} ~h^{~ k}_{j)~,vv}= 0  .
\end{align}
Using the previous results, we can show that the characteristic equation for the polarisation (A) is:
\begin{align}
 \xi_v \left( \xi_u + \left(\frac{\Rb}{b }\right)^{D-4} [  1 + 2\frac{ \lgb \ell_{\!\mathsmaller{2}}^2 (D-4)^2 }{ b^{2}}]\xi_v \right)  = 0
\end{align}
and for the polarisation (B):
\begin{align}
 \xi_v \left( \xi_u + \left(\frac{\Rb}{b }\right)^{D-4} [  1 - 4\frac{ \lgb \ell_{\!\mathsmaller{2}}^2 (D-4) }{ b^{2}}] \xi_v \right)  = 0
\end{align}
The solutions correspond to two characteristic hypersurfaces:
\begin{equation}
\xi_v  = 0, \qquad  \xi_u = - \hb \left[ 1 - 4 \displaystyle \gamma_{(A)/(B)} \frac{\lgb \ell_{\!\mathsmaller{2}}^2 (D-4) }{b^2} \right] \xi_v ,
\end{equation}
where
\begin{align}
\gamma_{(A)} = \frac{4-D}{2} ~~~~ \text{and} ~~~~ \gamma_{(B)} = 1  .
\end{align}
The second solution is inside the flat spacetime light-cone when $\xi_u$ is negative. This is the case if $\ltwo = 0$ (Einstein gravity) since $\Rb$ is a positive constant. For  $ b \sim  \sqrt{|\lgb|} \ell_{\!\mathsmaller{2}}$ we can always adjust $\xi_u$ to be positive for either the (A) or (B) polarisations depending on the sign of $\lgb$. This shows that Einstein-Gauss-Bonnet gravity allows for swift propagation regardless of sign of the Gauss-Bonnet coefficient.
If we would have considered instead the term $\hb_,^{\phantom{,} i j k l} h_{ k l , vv}$, which is produced by an action of the form $R - \frac{\lambda_3 \ell_{\!\mathsmaller{3}}^4}{8}R  R_{abcd}  R^{abcd}$ (where $\lambda_3$ stands for the sign of this correction), we would have obtained the characteristics:
\begin{equation}
  \xi_v  = 0 , \qquad
  \xi_u = - \displaystyle \hb \left[ 1 - \tilde{\gamma}_{(A)/(B)} \frac{ (D-4)(D-2) }{b^4}\lambda_3 \ell_{\!\mathsmaller{3}}^4 \right] \xi_v ,
\end{equation}
where
\begin{align}
\tilde{\gamma}_{(A)} = D-1  \qquad \text{and} \qquad \tilde{\gamma}_{(B)} = -1  \ .
\end{align}
These two examples can be matched to the two types of extended gravity amplitudes considered in \cite{camanho_causality_2014} (see their Eq. (3.17)).

Finally, returning to Table~\ref{table:resR3}, the only term that we have not discussed so far (and is not associated with a modification of the background) is $\mathcal{T}_{{}_{1}}^{ij} = \hb_{,}^{ ~  k (i } ~ \ld \d_v^2 h_{ k }^{\phantom{k} j )}$ of equation \eqref{eq:defT1}. This is simply the derivative of one of the terms that we have already studied. However, since it has four derivatives acting on the probe graviton it dominates the characteristic equations, which are determined from the principal part of the characteristic polynomial (as already mentioned in subsection \ref{subsec:methCar}). In this case the characteristic hypersurfaces do not depend on  the polarisation and are given by:
\begin{equation}
  \xi_v  = 0 , \qquad
  \xi_u  = -\xi_v \hb  = - \xi_v \displaystyle \left(\frac{\Rb}{b }\right)^{D-4} ,
\end{equation}
so that we recover the usual Shapiro time delay. Such a theory with equations of motion which are fourth order derivatives theory can suffer however from Ostrogradsky instabilities. We summarise our results in Table~\ref{table:sumup} for an independent basis of actions of the forms
$[R^2]$, $[R^3]$ and
$[R\nabla^2 R]$
 sorting them according to the four following categories:
\begin{itemize}[label=--]
\item Category 1 contains terms which do not modify the Hilbert-Einstein equations for the background and for the probe graviton. In this category, we have for instance $f(R)$ gravity and the Lovelock terms of order higher than two.
\item Category 2 contains terms that do not modify the background equations, but do modify the equations for the probe graviton in a way that leads to swift propagation. These are specified up to addition of terms from category 1.
\item Category 3 contains terms that do not modify the background equations, but their equations for the probe graviton are fourth-order in derivatives with characteristic hypersurfaces similar to Einstein-Hilbert gravity. These are specified up to addition of terms from categories 1 and 2.
\item Category 4 contains terms which have modified background equations, implying that we cannot apply our analysis in this case.  These are specified up to addition of terms from categories 1,2 and 3.
\end{itemize}

\begin{table}[!ht]

\begin{center}
\begin{tabular}{ | c | c |  }
    \hline
    \rule{0pt}{3ex}
\multirow{6}{6cm}{ Category 1 (no contribution) }  & $R^2 ,~ R^3 ,~R  \nabla_c \nabla^c R $   \\[0.4em]
&  $R_{a}^{~c}R^{ab}R_{bc} ~,~ R R^{ab}R_{ab}$   \\[0.4em]
&
$ ~ R^{ab}  \nabla_c \nabla^c R_{ab} $   \\[0.4em]
& $R  R_{abcd}  R^{abcd} - 4 R^{a b}   R_{a}^{~cde}   R_{bcde} + \frac{2}{3} R_{ab}^{ ~~ef}   R^{abcd}   R_{cdef}$ \\[0.4em]
& $R  R_{abcd}  R^{abcd} + \frac{8}{3} R_{a ~ c ~}^{~ e~ f}R^{a b c d}   R_{bedf}$ \\[0.4em]
\hline
\rule{0pt}{3ex}
\multirow{2}{6cm}{Category 2 (swift propagation) }   & $R  R_{abcd}  R^{abcd}$ \\[0.4em]
& $ R_{abcd}   R^{abcd}  - 4 R_{ab}   R^{ab} $ \\[0.4em]
\hline
\rule{0pt}{3ex}
\multirow{2}{6cm}{ Category 3 (higher derivative operators, infraluminal) }  & \multirow{2}{3 cm}{ $R_{ab}^{ ~~ef}   R^{abcd}   R_{cdef}$} \\[0.4em]
& \\
\hline
\rule{0pt}{3ex}
\multirow{2}{6cm}{Category 4 (background modified)}   & $R^{ab}  \nabla_c \nabla^c R_{ab} $ \\[0.4em]
& $R^{ab}  R_{ab} $ \\[0.4em]
\hline
 \end{tabular}
 \caption{ Basis of $[R^2]$, $[R^3]$ and
$[R\nabla^2 R]$ action terms, sorted by categories as detailed in the text. Our analysis is relevant as long as the background equations are not modified, i.e., for the terms in categories 1,2 and 3.}
\label{table:sumup}
\end{center}
\end{table}


\subsection{Swift Behaviour of $R^4$ Actions}
\label{sec:fourthorder}

As can be read from Table~\ref{table:R4}, the equations of motion of $[R^4]$ actions are linear combinations of the following three contributions:
\begin{equation}
\begin{split}
 \Vz {}^i {}^j &\equiv \hb_,^{\phantom{,} k l } \hb_{, k l } h^{ ij}_{\phantom{ij)} , v^4}, \\
\Uz{}^i {}^j &\equiv \hb_,^{\phantom{,} k l } \hb_{, k}^{\phantom{, k} (i } h^{j)}_{\phantom{j))} l , v^4}, \\
\Wz {}^i {}^j &\equiv \hb_,^{\phantom{,} i k}\hb_,^{\phantom{,} j l} h_{ k l , v^4} ,
\end{split}
\end{equation}
that lead to the following contributions to the characteristic equations. For $\Vz^{ij}$ we have:
 \begin{align}
\ca \left[\Vz^{ij} \,\right] = \cb [\Vz^{ij} \, ] = \left(\frac{\Rb}{b }\right)^{2(D-4)}\frac{   (D-2)(D-3)(D-4)^2}{ b^4} \xi_v^4  \, \ep_{ij} \ ,
\end{align}
which does not depend on the polarisation. For $\Uz^{ij}$ we have:
\begin{align}
\begin{split}
\ca \left[\Uz^{ij}\, \right] &= \left(\frac{\Rb}{b }\right)^{2(D-4)} \frac{   (D-4)^2}{ b^{4}}\left[1 + \frac{(D-2)(D-4)}{2}\right] \xi_v^4 \, \ep_{ij}, \\
\cb \left[\Uz^{ij} \, \right] &= \left(\frac{\Rb}{b }\right)^{2(D-4)}\frac{   (D-4)^2}{ b^{4}} \xi_v^4 \, \ep_{ij}\ ,
\end{split}
\end{align}
where the contributions of the two polarisations behave differently, but have the same sign, so that in general the sign cannot be adjusted by picking the appropriate polarisation. Finally, for $\Wz {}^i {}^j ~\equiv \hb_,^{\phantom{,} i k}\hb_,^{\phantom{,} j l} h_{ k l , v^4}$ we get:
\begin{align}
\begin{split}
\ca \left[ \Wz^{ij} \,\right]&=  - \left(\frac{\Rb}{b }\right)^{2(D-4)}\frac{   (D-4)^2}{ b^{4}} (D-3) \xi_v^4 \, \ep_{ij}, \\
\cb \left[ \Wz^{ij} \,\right] &= \left(\frac{\Rb}{b }\right)^{2(D-4)} \frac{  (D-4)^2}{ b^{4}} \xi_v^4  \, \ep_{ij} \ .
\end{split}
\end{align}
For this term we can adjust the sign by choosing an appropriate polarisation.

None of these three terms $\Vz, \Uz, \Wz$  can appear alone in the equations of motion. This is because they have fourth order derivatives of the form $\del_v^4$ acting on the probe graviton. The characteristic equations are dominated by  these  $\xi_v^4$ contributions, and if no other four derivative terms are present, the causal cone is locally degenerate.  We discuss this situation in detail in appendix \ref{app:causality} and show that in this case the full equations of motion cannot be hyperbolic with respect to the time direction when expanded around our background. The initial value problem is therefore not well-posed for gravitons propagating in the middle between the two beams.

Consider for instance the action:
\begin{align}
 \scr{L}_{ss4} = \frac{\sqrt{-g}}{16 \pi \ell_P^{D-2}} \left[ R +   \frac{\zeta(3)\lfour}{8} R^{abcd} \left(
R_{a ~ c }^{~ e~ f} R_{b~e}^{~h~g} R_{dhfg} - \frac{ R_{a b }^{~ ~ ef} R_{ce}^{~~hg} R_{dfhg}  }{4} \right) \right]  ,
\end{align}
present in the effective action originating from superstring theory as presented by\cite{tseytlin_r**4_2000}.\footnote{It differs from the expression of \cite{gross_superstring_1986} by factors of $\L_{\!\mathsmaller{4}}$, the fourth order Lovelock Lagrangian. Such factors are irrelevant in our analysis since they do not influence the equations of motion for the probe graviton on our background.}
Using the results of section \ref{sec:EOM}, the equations of motion read:
\begin{align}
 2h^{ij}{}_{,uv} + 2\hb h^{ij}{}_{,v^2}    - \lfour \frac{ \zeta(3)}{2} \hb_{,}^{~kl} \hb_{,kl} h^{ij}_{~~ , v^4} =  0  \ ,
\end{align}
and have a degenerate causal cone  as discussed above. However, this can be fixed by adding extra, lower order terms to the action which have equations of motion with four derivatives acting on the probe graviton.
Table \ref{table:resR3} reveals two such terms, $R^{a b}   R_{a}^{~cde}   R_{bcde}$ and $R_{ab}^{ ~~ef}   R^{abcd}   R_{cdef}$, which do not modify the background equations. 
We therefore modify our action to:
\begin{align}
 \scr{L}_{mod} = \frac{\sqrt{-g}}{16 \pi \ell_P^{D-2}}    &\left[R + \lthree R^{ab}   R_{a cde}   R_{b}^{\phantom{b} cde} \right. \\ \nn
& \left. + \lfour \frac{\zeta(3)}{8} R^{abcd} \left(
R_{a ~ c }^{~ e~ f} R_{b~e}^{~h~g} R_{dhfg} - \frac{1}{4} R_{a b }^{~ ~ ef} R_{ce}^{~~hg} R_{dfhg}   \right) \right] \nn
\end{align}
which reduces to the previous one in the limit $ \ell_{\!\mathsmaller{3}} \rightarrow 0 $.
We can use our previous results to find that the characteristic equations have the following solutions:
\begin{equation}
  \xi_v  = 0 , \qquad
  \xi_u = - \hb \displaystyle \left[ 1 - \gamma_{(A)/(B)} \frac{\zeta(3) \lfour}{16 b^2 \lthree}(D-2)(D-3) \right]  \xi_v \ ,
  \end{equation}
where
\begin{align}
 \gamma_{(A)} = -1 \text{ ~~~ and ~~~  } \gamma_{(B)} = \frac{D-4}{2} \ .
\end{align}
Since the sign of $\gamma$ can be chosen to be either positive or negative, we have swift propagation if:
\begin{align}
 \frac{\lthree b^2}{\lfour} < \frac{\zeta(3)}{8} (D-2)(D-3) |\gamma|
\end{align}
therefore when varying $\frac{\lthree b^2}{\lfour}$ from infinity (no fourth order term) to zero (only fourth order term), the theory starts  infraluminal, then admits swift propagation and finally the causal cone collapses and becomes degenerate.

Below a critical value:
\begin{align}
 b_c \sim \hb \sqrt{\left|\frac{\lfour}{\lthree}\right|}
\end{align}
the local causal cone envelops a region including slices of constant time.

We close this section by supplementing the unitarity and analyticity  constraints of  \cite{bellazzini_quantum_2015} with some constraints implied by requiring the absence of
a degenerate causal cone in actions composed of the Einstein-Hilbert action supplemented by quartic corrections only. We then obtain two algebraic relations between the coefficients $c_i$ of~\cite{bellazzini_quantum_2015} that are complementary to their positivity requirements. These are:
\begin{equation}
\label{eq:causalconstr}
\begin{split}
 2(c_7+2c_2+4c_4) &= - 2(D-3) (2c_6+c_7) ,\\
2(c_6 + 2c_2 + 8c_3) &= -(D-3)(D-4) (2c_6+c_7)  .
\end{split}
\end{equation}


%

\section{Four Dimensional Case}
\label{sec:4D}

In this section we collect all the results regarding swift propagation in $D=4$. There are several differences from the cases with $D>4$. Solving the Einstein equation~\eqref{EoMbkd} in four dimensions gives:
\begin{align}
\label{hbkd4D}
 \hb(x_i) =  \begin{cases}
\displaystyle 8 \pi \rho \ell_P^2 r_0^2    \log \frac{\Lambda}{r} & \qquad r>r_0
 \\[1.2em]
\displaystyle ~ 4 \pi \rho  \ell_P^2 r_0^2 \left[1+ 2 \log \frac{\Lambda}{r_0}- \frac{r^2}{r_0^2}  \right] & \qquad r\leq r_0
\end{cases}
  \ .
\end{align}
Where $\Lambda$ is a positive constant which is not fixed by the Einstein equations. Following the discussion of Section~\ref{subsec:timeshift}, we view this as an IR cutoff to our theory related to the length of the beams, indicating that we should take $r< \Lambda$.

Defining the constant $a$ by
\begin{align}
 a \equiv  8 \pi \rho  \ell_P^2 r_0^2  > 0
\end{align}
we obtain
\begin{equation}
\label{eq:deriv4D}
\begin{split}
\hb (\vec b) = & \, a \log \frac{\Lambda}{r} \\
 \d_i \d_j \hb (\vec{b}) =  & \, -  \frac{a}{b^4} (b^2 \delta_{ij} - 2b_i b_j)
 \\
 \d_i \d_j \d_k \d_l \hb (\vec{b})  = & \, \frac{2a}{b^{8}}  \left[ - 4 b^2  ( \delta_{ij} b_k b_l  + \dots )
 + b^4 (\delta_{kl} \delta_{ij} + \dots )
 \right.
 \\
 &
~~~~~~~~~~~~~~~~~~~~~~~~ \left.
 + 24 b_i b_j b_k b_l \right]  ,
\end{split}
\end{equation}
where the $\dots$ stand for all index permutations of a given tensor structure.

The results of Tables \ref{table:resR2}, \ref{table:resR3} and \ref{table:R4} are left unchanged (since we have not used the particular form of $\hb$ there). However the study of swiftness is modified  by the fact that only the class (A) polarisation is compatible with the two transverse dimensions left in $D=4$.  This implies that we cannot choose the nature of the time shift by going from type (A) to type (B) polarisations. The presence or absence of a swift propagation therefore depends on the particular sign of the coupling constants of the extended gravity theory.

Next, we compute the characteristic equations for the various contributions to the equations of motion (suppressing the $\epsilon^{ij}$):
\begin{equation}
\label{charact4D}
\begin{split}
& \ca \left[ -\frac{1}{2} \ld h^{ij} \right] =
  2\xi_u  \xi_v    + 2 \xi_v^2 a \log \frac{\Lambda}{r}  ,
\\ 
& \ca \left[ \hb_,^{\phantom{,} k (i} h^{j)}_{\phantom{j)} k , vv} \right] =  0,
\\ 
& \ca \left[\hb_{,}^{ ~ k (j } ~ \ld \d_v^2 h^{i)  }_{~ k } \right] = 0,
\\ 
& \ca \left[ \hb_{,}^{ ~ i j k l } ~ \d_v^2 h_{ k l }  \right] =  -\frac{12a}{b^4} \xi_v^2
\\ 
& \ca \left[ \hb_{,}^{~lk} \hb_{, k}^{~~ (i} h^{j )}_{~~ l , v^4} \right] = \frac{a^2}{b^4} \xi_v^4
\\ 
& \ca \left[ \hb_{,}^{~ik} \hb_{,}^{~jl} h_{kl , v^4} \right] = - \frac{a^2}{b^4} \xi_v^4
\\ 
& \ca \left[ \hb_{,}^{~kl} \hb_{,kl} h^{ij}_{~~ , v^4} \right] = \frac{2 a^2}{b^4} \xi_v^4
\end{split}
\end{equation}
Using the results of Table \ref{table:resR2} one can easily check that the Gauss-Bonnet combination gives a vanishing contribution in $D=4$. This is expected since the Euler density is topological in  $D=4$. Furthermore, we see that the only non-vanishing contribution from $[R^3]$ terms is $ \hb_{,}^{ ~ i j k l } ~ \d_v^2 h_{ k l } $, which is present in the equations of motion derived from $R  R_{abcd}  R^{abcd}$, $R^{a b}   R_{a}^{~cde}   R_{bcde}$ and $R_{a ~ c ~}^{~ e~ f}R^{a b c d}   R_{bedf}$. The most general Lagrangian build from $[R^2]$ and $[R^3]$ terms for which our background~\eqref{hbkd4D} is a solution is then of the form:
\begin{align*}
\L = \frac{\sqrt{-g}}{16 \pi \ell_P^{2}}    &\left[R + \lthree ( d_{1} R  R_{abcd}  R^{abcd} R^{a b} + d_{2}   R_{a}^{~cde}   R_{bcde}  + d_{3} R_{a ~ c ~}^{~ e~ f}R^{a b c d}   R_{bedf}) + (\dots) \right]  \ ,
\end{align*}
where the $(\dots)$ includes all the $[R^2]$ and $[R^3]$ terms with vanishing contributions to the characteristic equations (see  Eq.~\eqref{charact4D} and Table~\ref{table:resR3}). We obtain the characteristic equation:
\begin{align}
\label{eq:charact4D}
 \xi_u \xi_v + \xi_v^2 a \left[ \log \frac{\Lambda}{b}+ \frac{6 \lthree}{b^4}(8d_{1} + 2d_{2}+3d_{3}) \right] = 0\ .
\end{align}
The presence of the logarithm and how to deal with it has been discussed at the end of Section~\ref{subsec:timeshift}. The presence of swift propagation for $b \sim \ell_3$ then depends on the sign of the last term in Eq.~\eqref{eq:charact4D}. We observe swift propagation if:
\begin{align}
8d_{1} + 2d_{2}+3d_{3} < 0 \ .
\end{align}
Since we are in four dimension, we have redundancy between the different operators listed in Table~\ref{table:resR3}. For instance, the fact that the third order Lovelock theory~\eqref{eq:lov3} is topological implies that we can trade one of the $d_i$ for the two others.

Finally, turning to the $[R^4]$ terms, we see that we get once more a degenerate causal cone. Notice that it is not possible to fix this behaviour by adding $[R^3]$ terms as we did in Section~\ref{sec:fourthorder}.

\section{Conclusions}

In this work we presented a simple, albeit not exhaustive, way of studying the local causal structure of a large variety of extended gravity theories. By considering perturbations propagating in parallel to two beams described by the stress-energy tensor~\eqref{Tmunu}, we found that many extended gravity theories allow for swift propagation. The background beams were chosen to mimic the successive scattering events of gravitons that~\cite{camanho_causality_2014} used to find a sizeable time-shift. In this sense, we recover and extend the results of~\cite{camanho_causality_2014} in an alternative fashion.  We have further studied 
all possible theories based on $[R^2]$, $[R^3]$, $[R\nabla^2R]$ and $[R^4]$ terms around our background~\eqref{hbkd} and found several interesting properties:
\begin{itemize}
 \item Swiftness appears to be a quite general feature. Indeed we have shown that a large class of $[R^2]$,  $[R\nabla^2R]$ and $[R^3]$ theories exhibit swift graviton propagation.
 \item $f(R)$ theories and Lovelock theories, known to be free of Ostrogradsky instabilities, have the same local causal structure as general relativity around our background. Therefore, gravitons remain infraluminal in these theories. The case of Gauss-Bonnet gravity is an exception.
\item Several theories, including those with certain $[R^4]$ terms in their action, have equations of motion with four derivatives acting on the probe graviton. This is not a game-ending property from the point of view of causality. However, we showed that such theories often lead to a degenerate causal cone and argued that this situation is related to non hyperbolicity of the equations of motion. We have presented an easy way of curing this pathological case in $D>4$  by adding carefully chosen $[R^3]$ terms.
\item In $D>4$, the signs of the coupling constants of the extended gravity theories were irrelevant when dealing with swift propagation. This was not the case however, for $D=4$.
\end{itemize}

Our results can be used to constrain the possible extended gravity theories, but only if they do not arise from an effective theory with scale $l_e$. Indeed, swift behaviour requires impact parameter $b \sim l_e$, precisely the scale at which the effective theory approach should be replaced by the full UV theory. For a particular case of string theory, it was argued in~\cite{dappollonio_regge_2015} that superluminal propagation can still occur as long as one does not consider the full infinite tower of effective operators. An argument which can be related to the observation of~\cite{camanho_causality_2014} that considering the effect of a finite tower of higher spin particles would not solve the time advance issue.

It would be interesting to check if our background could be obtained from a well-defined initial value problem by studying the full hyperbolicity of our equations of motion. This will require to study more general perturbations around $\hb$. Finally, higher derivatives theories are also subject to more constraints from unitarity and positivity. Understanding the interplay between causality constraints, swiftness and unitarity will also be an important step.

\section*{Acknowledgements}
We would like to thank Luis Lehner, Marco Meineri and Giuseppe Policastro for valuable comments.
Y.O. would like to thank the LPTHE and ILP for their hospitality. This work  was supported in part by French state funds
managed by the ANR within the Investissements d'Avenir program (reference
number ANR-11-IDEX-0004-02) and in part by the ERC Higgs@LHC.
The work of Y.O.  is supported in part by the I-CORE program of Planning and Budgeting Committee (grant number 1937/12), and by the US-Israel Binational Science Foundation.
Research at Perimeter Institute is
supported by the Government of Canada through Industry Canada and by the Province
of Ontario through the Ministry of Research \& Innovation.

\appendix
\addcontentsline{toc}{section}{Appendices}
\section*{Appendices}

\section{List of Metric Components, Christoffel Symbols and Riemann Components}\label{app:List}
We denote $x^\mu = (u, v, \vec x)$. The  zeroth order metric components are given by:
\begin{equation}
g_{\mu\nu}^{(0)} =
\left(
	\begin{array}{rrr}
 	  h_0(\vec x) & -\frac{1}{2} &   \\
      -\frac{1}{2} & 0 &  \\
      & & \delta_{ij}\\
	\end{array}
\right) ,
\qquad
g^{(0)\mu\nu} =
\left(
	\begin{array}{rrr}
 	  0 & -2~~ &   \\
      -2 & -4h_0(\vec x) &  \\
      & & \delta^{ij}\\
	\end{array}
\right) , \qquad
\end{equation}
where $h_0(\vec x)$ is given by Eq. \eqref{hbkd}.
As a rule of thumb at zeroth order in the probe graviton contribution:
\begin{equation}
\begin{split}
&V^{u} = -2 V_{v}, \qquad
V^{v} = -2V_{u} -4 h_0 V_{v},\\
&V_{u} = h_0 V^{u} -\frac{1}{2} V^{v}, \qquad
V_{v} = -\frac{1}{2}V^{u},
\end{split}
\end{equation}
and so, a vector that only has a lower $u$ component will only have an upper $v$ component and the other way around.

\subsection{Zeroth Order in the Probe Graviton Contribution}
The non-vanishing Christoffel symbols are given by:\footnote{The others are implied by symmetry or vanish.}
\begin{equation}\label{eq:0_order_christoffels}
\begin{split}
&\Gamma_{iuu}^{(0)}	 = -\frac{1}{2} \del_i h_0, \qquad
\Gamma_{uiu}^{(0)} = \frac{1}{2} \del_i h_0,\\
&\Gamma^{i(0)}_{uu} = -\frac{1}{2} \del_i h_0, \qquad
\Gamma^{v(0)}_{iu} = - \del_i h_0.
\end{split}
\end{equation}
Those vanish when summing right in the middle between two beams.
The non-vanishing Riemann components are:
\begin{equation}
\begin{split}
&R^{(0)}_{iuku} = -\frac{1}{2} \del_k\del_i h_0,\\
&R^{i}{}_{uku} = -\frac{1}{2} \del_k \del^i h_0, \qquad
R^{v}{}_{iku} = - \del_k \del_i h_0, \qquad
\\
&R^{(0)ivkv} = -2 \del^k \del^i h_0.
\end{split}
\end{equation}
The non vanishing Ricci tensor components are:
\begin{equation}
R^{(0)}_{uu} = -\frac{1}{2} \del_i \del^i h_0, \qquad R^{(0)vv} = - 2 \del_i \del^i h_0,
\end{equation}
those vanish outside the beam.
And the Ricci scalar vanishes identically:
\begin{equation}
R^{(0)} = 0.
\end{equation}

\subsection{First Order in the Probe Graviton Contribution}
We allow for a probe graviton of the form:
\begin{equation}
\delta g^{(1)}_{ij} = h_{ij} (u,v),
\qquad
\delta g^{(1)ij} = -h_{ij} (u,v).
\end{equation}
Which leads to the following non-vanishing Christoffel symbols:
\begin{equation}
\begin{split}
\Gamma^{(1)}_{uij} = -\frac{1}{2} \del_u h_{ij},
\qquad
\Gamma^{(1)}_{iuj} = \frac{1}{2} \del_u h_{ij},\\
\Gamma^{(1)}_{vij} = -\frac{1}{2} \del_v h_{ij}, \qquad
\Gamma^{(1)}_{ivj} = \frac{1}{2} \del_v h_{ij},
\end{split}
\end{equation}
or alternatively:
\begin{equation}
\begin{split}
&\Gamma^{i(1)}_{uu} = \frac{1}{2} h^{ij} \del_j h_0,
\qquad
\Gamma^{i(1)}_{uj} = \frac{1}{2} \del_u h_{ij},
\\
&\Gamma^{v(1)}_{ij} =  \del_u h_{ij} +2h_0 \del_v h_{ij},
\qquad
~~~\Gamma^{i(1)}_{vj} = \frac{1}{2} \del_v h_{ij},
\qquad
~\Gamma^{u(1)}_{ij} =  \del_v h_{ij}.
\end{split}
\end{equation}
The non vanishing components of the Riemann tensor are given by:
\begin{equation}
\begin{split}
&R^u{}^{(1)}_{iju} = -\del_u \del_v h_{ij},
\qquad
R^u{}^{(1)}_{ijv} = -\del_v^2 h_{ij},
\qquad
R^u{}^{(1)}_{uiu} = -\frac{1}{2}\del_k h_0 \, \del_v h_{ik},
\\
&R^v{}^{(1)}_{kij} = \del_i h_0 \del_v h_{kj} -\del_j h_0 \del_v h_{ki},
\qquad~~~~~~~~~~
R^v{}^{(1)}_{uuj} = h_0 \del_i h_0 \del_v h_{ij},
\\
&R^v{}^{(1)}_{kuj} = \del_u^2 h_{kj}+2 h_0 \del_u \del_v h_{kj},
\qquad ~~~~~~~~~~~~
R^v{}^{(1)}_{kvj} = \del_v \del_u h_{ij} +2h_0 \del_v^2 h_{ij},
\\
&R^v{}^{(1)}_{iuv} = -\frac{1}{2} \del_k h_0 \del_v h_{ik},
\qquad~~~~~~~~~~~~~~~~~~~~~
R^v{}^{(1)}_{vuj} = -\frac{1}{2} \del_i h_0 \del_v h_{ij},
\\
&R^i{}^{(1)}_{ujk} = -\frac{1}{2} \del_v h_{ij} \del_k h_0 +\frac{1}{2} \del_v h_{ik} \del_j h_0,
\\
& R^i{}^{(1)}_{uuk} = \frac{1}{2} \del_u^2 h_{ik} - \frac{1}{2} h^{ij} \del_k\del_j h_0,
\qquad~~~~~~~~~~~~~~
 R^i{}^{(1)}_{vuk} = \frac{1}{2}\del_u \del_v  h_{ik},
\\
& R^i{}^{(1)}_{juk} = \frac{1}{2} \del_v h_{ik} \del_j h_0 -\frac{1}{2} \del_v h_{jk}\del_i h_0,
\qquad~~~~~~~~
R^i{}^{(1)}_{uvk} = \frac{1}{2} \del_u \del_v h_{ik},
\\
& R^i{}^{(1)}_{vvk} = \frac{1}{2} \del_v^2 h_{ik},
\qquad~~~~~~~~~~~~~~~~~~~~~~~~~~~~~~~~
R^i{}^{(1)}_{uuv} = -\frac{1}{4} \del_v h^{ij} \del_j h_0,
\end{split}
\end{equation}
or alternatively:
\begin{equation}
\begin{split}
&R^{(1)}_{ukij} = \frac{1}{2}( \del_j h_0 \del_v h_{ki} - \del_i h_0 \del_v h_{kj}),
\qquad
R^{(1)}_{uvui} = \frac{1}{4} \del_j h_0 \del_v h_{ij},
\\
&
R^{(1)}_{ukui} = -\frac{1}{2} \del_u^2 h_{ki},
\qquad
R^{(1)}_{ukvi} = -\frac{1}{2} \del_v \del_u h_{ki},
\qquad
R^{(1)}_{vijv} = \frac{1}{2} \del_v^2 h_{ij}.
\end{split}
\end{equation}
The non vanishing Ricci tensor components are given by:
\begin{equation}
\begin{split}
& R^{(1)}_{ij} = 2\del_u\del_v h_{ij} +2 h_0 \del_v^2 h_{ij},
\qquad
R^{(1)}_{ui} = \del_v h_{ij} \del_j h_0 -\frac{1}{2} \del_i h_0 \del_v h_k^k,
\\
& R^{(1)}_{uu} = \frac{1}{2} h^{kj} \del_k \del_j h_0 -\frac{1}{2} \del_u^2 h_k^k,
\qquad
 R_{uv} = -\frac{1}{2} \del_u\del_v h_k^k,
\qquad
R_{vv} = -\frac{1}{2} \del_v^2 h_k^k.
\end{split}
\end{equation}
If we assume that the perturbation is traceless we are left with:
\begin{equation}
 R^{(1)}_{ij} = 2\del_u\del_v h_{ij} +2 h_0 \del_v^2 h_{ij},
\qquad
R^{(1)}_{ui} = \del_v h_{ij} \del_j h_0,
\qquad
 R^{(1)}_{uu} = \frac{1}{2} h^{kj} \del_k \del_j h_0.
\end{equation}
Note that there are no lower $v$ indices and hence no upper $u$ indices.
The Ricci scalar reads:
\begin{equation}
R^{(1)} = 4 \del_u \del_v h_k^k +4 h_0 \del_v^2 h_k^k,
\end{equation}
and vanishes for a traceless perturbation:
\begin{equation}
R^{(1)}=0.
\end{equation}

\section{Characteristic Method for a Scalar Field}
\label{app:charact}
In this appendix we explain the idea behind the method of characteristics for the simplest example of a scalar field.

 Assume that the polynomial corresponding to the principal  part of the equation of motion takes the form:
\begin{equation}\label{Principle}
P_m(\xi_\alpha) = \xi_1^m + \xi_1 \xi_2^{m-1} +\dots = 0.
\end{equation}
and $\alpha=(1,\dots,n)$, and assume that it admits a non-trivial solution $\xi_{0\alpha}$ for which the first non-vanishing component is $\alpha=k$. Define
\begin{equation}
\tilde \xi_\alpha =
\begin{cases}
\xi_\alpha
\qquad\qquad\qquad~~
 \alpha \leq k
\\
\xi_\alpha - \left(\frac{\xi_{0\alpha}}{\xi_{0k}}\right)\xi_k
\qquad \alpha > k.
\end{cases}
\end{equation}
This corresponds to the coordinate transformation:
\begin{equation}
\tilde x_\alpha =
\begin{cases}
x_\alpha
\qquad\qquad\qquad\qquad~~~~~
 \alpha \neq k
\\
x_k + \sum\limits_{\beta=k+1}^{m}\left(\frac{\xi_{0\beta}}{\xi_{0k}}\right)x_\beta
\qquad \alpha = k.
\end{cases}
\end{equation}
of the original equation.
In these coordinates the characteristic polynomial admits a solution of the form $\tilde\xi_{0\alpha} = \delta_\alpha^k$ hence forcing the coefficient of $(\tilde \xi_k)^m$ in \eqref{Principle} to vanish identically. In this case the equation of motion loses its predictive power for $(\tilde \del_k)^m$ and can no longer predict its value based on the set of $m-1$ derivatives in all directions and all $m$ mixed derivatives except for $(\tilde \del_k)^m$. This defines the characteristic hypersurface as the hypersurface of constant $\tilde x_k$, or equivalently the hypersurface orthogonal to $\xi_{0\alpha}$ which is the hypersurface on which a full set of Cauchy data does not fix the external evolution.

For instance consider the equation:
\begin{equation}\label{diff_app_B}
\del_u\del_v \phi +a \del_u\del_v^3 \phi +b \del_v^4\phi = 0.
\end{equation}
The characteristic polynomial is:
\begin{equation}
P_4(\xi_u,\xi_v) = a \xi_u \xi_v^3 +b \xi_v^4
\end{equation}
which vanishes for $\xi_v = 0$ or $\xi_u = -\frac{b}{a} \xi_v$. This means that the characteristic hypersurfaces are orthogonal to $(u,v) = (1,0)$ and $(u,v) = (-b/a,1)$ respectively. In terms of the
new coordinates $(\tilde u, \tilde v) = (u-\frac{a}{b}v,v)$ \eqref{diff_app_B} does not have a $\tilde\del_u^4$ term hence does not offer predictions in this direction.

\section{Causality and Hyperbolicity}
\label{app:causality}

In this appendix we summarise some useful notions regarding causality and hyperbolicity. This includes a formal definition of hyperbolicity and its relation to well-posed initial value problems, the definition of the causal cone and the notion of causality  \cite{courant_methods_1962,Hormander:1983}. We also quote a theorem which in the context of this paper can be used to demonstrate that when the projection of the  causal cone to the $(u,v)$ plane is degenerate the full equations of motion are not hyperbolic.

Let us start by defining hyperbolicity. Let $P(\d)$ be a differential operator of order $m$: $P(\d) = \sum_{i=0}^m P_i(\d)$, where all terms of order $i$ are regrouped in $P_i$. The highest order terms, $P_m(\d)$ are called the principal part. The operator $P(\partial)$ is said to be hyperbolic with respect to a vector $\xi$ if all the following conditions are satisfied:
\begin{enumerate}
 \item $\xi$ is not a root of the principal part: $P_m(\xi) \neq 0$.
 \item $P_m(V  + \tau \xi) = 0$ has only real roots in $\tau$ where $V$ is a real vector. We furthermore define the cone $\Gamma(P,\xi)$ by $V \in \Gamma(P,\xi)$ if and only if all the previous roots are strictly negative.
 \item $P_m$ is stronger than the remaining terms in $P$ in the sense that there exists $C \in \mathbb{R}$ positive such that $\forall i \in [0,m]$ and  $\forall \xi \in \mathbb{R}^D $
\begin{equation}
 \frac{\sum_{\alpha \geqslant 0 , i } | P_i^{(\alpha)}(\xi)|^2}{\sum_{\alpha \geqslant 0 } |P_m^{(\alpha)}(\xi)|^2} < C \ ,
\end{equation}
where we use the usual parenthesis notation for the  derivative of order $\alpha$ with respect to all possible parameters.
\end{enumerate}
Notice that the last condition is automatically satisfied if $P = P_m$, so that the principal part of $P$ is hyperbolic if it satisfies the two first conditions.

We also define the causal cone $\Gamma^0(P,\xi)$ more formally as follows. $\Gamma^0(P,\xi)$ is the set of all vectors inside $\Gamma(P,\xi)$. That is, every element $V$ in $\Gamma^0(P,\xi)$ is such that $V \cdot \theta \geqslant 	 0$ for all $\theta \in \Gamma(P,\xi)$. It can usually be obtained by considering the cone bounded by the characteristics in the direction of $\xi$ (in the future-time direction if $\xi$ is taken to be the time direction).

Suppose we fix the boundary conditions on a hyperspace $\scr{I}$ with normal $\xi$, if the initial conditions are non-zero only on a compact convex subset $K$ of $\scr{I}$, then causality is the  requirement that the solution to our Cauchy problem vanishes outside $K +  \Gamma^0(P_m,\xi)$. This requirement holds as long as $P$ is hyperbolic \cite{Hormander:1983}.

Next, we quote a theorem which applies to hyperbolic systems and which we find useful in arguing that our system is not hyperbolic in the case of a degenerate causal cone.
The theorem states that for a hyperbolic system $P = P_m + P_{m-1} +\dots$ with respect to a vector $\xi$, if $V$ is not proportional to $\xi$ and $\tau_0$ is a root of $P_m(V + \tau\xi)$ of degeneracy $\mu$ then  $V+\tau_0 \xi$ is a root of degeneracy $\mu-j$ of $P_{m-j}$, $j=0,\dots,m-1$.

To show that for the cases we study in this paper a system with a degenerate causal cone cannot be hyperbolic we use the following argument. Suppose we decompose the characteristics polynomial $P(\xi)$ as follows:
\begin{equation}
 P(\xi) = Q(\xi_u,\xi_v) +  R(\xi_u,\xi_v,\xi_i) \ ,
\end{equation}
where  $R(\xi_u,\xi_v,0)=0$. In this paper, we have studied the term $Q$, since we have restricted the motion of our probe graviton to the $(u,v)$ plane. For the case of a degenerate causal cone, we have shown that $Q$ takes the generic form:
\begin{equation}
\label{eq:polydecomp}
Q(\xi_u,\xi_v) =  \xi_u \xi_v  + a \xi_v^2 h+ b \xi_v^4  \ ,
\end{equation}
where $a, b$ are real non-zero constants. Suppose the full equations of motion are hyperbolic with respect to $\tilde{\xi} = (\xi_u,\xi_v,0,\dots)$, with $\xi_v \neq 0$ (this includes for instance the time direction $\xi_{time} \propto (1,1,0,\dots)$). Given $V=(V_u,V_v,0,\dots)$ not proportional to $\xi$ we have:
\begin{equation}
 P_m(V + \tau\xi) =  Q_m(V + \tau\xi) =  b (V_v + \tau \xi_v)^4 \ .
\end{equation}
We find that $\tau_0 = - V_v/\xi_v$ is a root of  $ P_m(V + \tau\xi)$ of degeneracy four.  The previous theorem then implies that $\tau_0$ must be a second order root of $ P_2(V + \tau\xi) =  Q_2(V + \tau\xi)$, if $P$ is hyperbolic. Since  $Q_2(\xi_u,\xi_v) =  \xi_u \xi_v  + a \xi_v^2 $ and $V$ is not proportional to $\xi$, $\tau_0$ is only a first order root, contradicting the result of the previous theorem. We therefore conclude that the full equations of motion are not hyperbolic with respect to $\xi$.
This proves that in the case of a degenerate causal cone (at least in the context of the cases studied in this paper), the full equations of motion are not hyperbolic with respect to the time direction when evaluated locally in between the two beams.

\section{Replacement Rules}
\label{app:rules}
In this appendix we detail some of the replacement rules which we used in \textit{Mathematica} in order to obtain our results.

\subsection{Zeroth Order Equations of Motion}

We consider a background of the form $\hb = \left(\frac{\Rb}{r }\right)^{D-4}$, as in equation \eqref{hbkd} for $r >r_0$, with $D>4$ and where $R_b$ was defined in \eqref{eq:DefRb} and is a positive constant. We denote by $\vec{b}$ the impact parameter, that is, the location in which we evaluate our equations of motion. The value and first derivatives of $\hb$ at this point are given by:
\begin{equation}
\begin{split}
\hb (\vec b) = & \, \left(\frac{\Rb}{b }\right)^{D-4},\\
 \d_i \hb (\vec{b}) = & \, -\left(\frac{\Rb}{b }\right)^{D-4} \frac{(D-4)}{b^{2}} b_i, \\
 \d_i \d_j \hb (\vec{b}) =  & \, - \left(\frac{\Rb}{b }\right)^{D-4} \frac{(D-4)}{b^{4}} (b^2 \delta_{ij} - (D-2)b_i b_j) ,
 \\
 \d_i \d_j \d_k \hb (\vec{b}) = & \, \left(\frac{\Rb}{b }\right)^{D-4} \frac{(D-4)(D-2)}{b^{6}} \left[ b^2 (\delta_{ij} b_k + \delta_{jk} b_i +\delta_{ik} b_j) \right.
 \\
 &
~~~~~~~~~~~~~~~~~~~~~~~~
 ~~~~~~~~~~~~~~~~~~~~~~~~~~~~~~~~
  \left. - D b_i b_j b_k \right] ,
 \\
 \d_i \d_j \d_k \d_l \hb (\vec{b})  = & \, \left(\frac{\Rb}{b }\right)^{D-4} \frac{(D-4)(D-2)}{b^{8}}  \left[ - b^2 D ( \delta_{ij} b_k b_l  + \dots )
 \right.
 \\
 &
~~~~~~~~~~~~~~~~~~~~~~~~ \left.
 + b^4 (\delta_{kl} \delta_{ij} + \dots )
 + D (D +2) b_i b_j b_k b_l \right]  ,
\end{split}
\end{equation}
where the $\dots$ stands for all index permutations of a given tensor structure.
At the midpoint between two identical beams we obtain:
\begin{equation}
\begin{split}\label{eq:h_0_derivatives}
 \d_i \hb (\vec{b}) & \rightarrow \d_i (\hb + \hbt) (\vec{b}) = \d_i (\hb(\vec{b}) + \hb(-\vec{b}) ) = 0 , \\
 \d_i \d_j \hb (\vec{b}) & \rightarrow \d_i \d_j (\hb  + \hbt )(\vec{b})  =  2 \d_i \d_j \hb (\vec{b}) ,\\
 \d_i \d_j \d_k \hb (\vec{b}) &\rightarrow \d_i \d_j \d_k (\hb  + \hbt )(\vec{b})  = 0, \\
\d_i \d_j \d_k \d_l \hb (\vec{b}) & \rightarrow \d_i \d_j \d_k \d_l (\hb  + \hbt )(\vec{b})  =  2 \d_i \d_j \d_k \d_l \hb (\vec{b})  .
\end{split}
\end{equation}

Before we start, we reiterate a number of useful facts. First, \eqref{eq:h_0_derivatives} implies that   the zeroth order Christoffel symbols \eqref{eq:0_order_christoffels} and any  even number of derivatives acting on them vanish when evaluated at the midpoint between the two beams. With a little abuse of notation we write:
\begin{equation}\label{chris_van}
\Gamma^{(0)}_{\mu\nu\rho} (\vec{b})=0 ,  \qquad \d_\alpha \d_\beta \Gamma^{(0)}_{\mu\nu\rho} (\vec{b}) = 0, \qquad \text{etc} .
\end{equation}
Similarly any odd number of derivatives acting on a zeroth order Riemann tensor vanishes at the midpoint:
\begin{equation}\label{Riem_van}
\d_{\mu} R^{(0)}_{\alpha\beta\gamma\delta} (\vec{b})=0, \qquad \text{etc}.
\end{equation}
Finally, notice that there are no zero order quantities with lower $v$ or upper $u$ indices.

In order to make sure that $\hb$ given in \eqref{hbkd} is indeed a valid background we have to make sure that it solves the equations of motion of the full extended gravity theory. A sufficient condition is that the zeroth order equations of motion take the form \eqref{EoMbkd}. This is the case for many of the theories studied in this paper. Since all zeroth order quantities have two lower $u$  indices  and no lower $v$  indices  and since $h_0$ does not depend on $u$ and $v$ we have at most one $h_0$ (and therefore, one Riemann or Ricci tensor) contribution to the $E_{uu}$ equation of motion and no contributions to the other components of the equations of motion. This is because we are left with lower $u$ indices  which cannot be contracted. However it is always possible to have contributions to the equations of motion of the form $\del^4 h_0$, $\del^6 h_0$ from additional space derivatives acting on the Riemann tensor. We can use the following replacement rules:
\begin{align}
&\nabla{}_\alpha \nabla{}^\alpha R{}^\mu {}^\nu = -\frac{1}{2} \vec{\d}^{4} \hb, \\
&\nabla{}^\alpha \nabla{}_\alpha \nabla{}^\beta \nabla{}_\beta R{}^\mu {}^\nu =  -\frac{1}{2} \vec{\d}^{6} \hb .
 \end{align}

For polynomials of third and fourth order in the Riemann or Ricci tensors we get that the zeroth order equations of motion are automatically satisfied by our background since they contain at least two instances of the Riemann or Ricci tensors.
The contributions to the zeroth order equations of motion for all independent gravity actions containing one or two Riemann tensors and at most two covariant derivatives are listed in Table \ref{table:EOM_0}.\footnote{We did not consider as independent, gravity actions which are  related by integration by parts, use of the second Bianchi identity or the addition of $R^3$ terms which do not contribute to the equations of motion at zeroth order in the probe graviton.}

\begin{table}[!ht]
\begin{center}
\begin{tabular}{ | c | c |}
   \hline
  Action & Background equations  \\
   \hline
   \rule{0pt}{3ex}
$  R$ &   $ -\frac{1}{2} \vec \del^2 h_0 $  \\[0.4em]
$  R^2$ &   $ 0$ \\[0.4em]
$  R_{ab}  R^{ab}$ &   $ -\frac{1}{2} \vec{\d}^4 \hb $ \\[0.4em]
$ R_{abcd}   R^{abcd} $  &  $ -2 \vec{\d}^4 \hb $  \\[0.4em]
$R  \nabla_c \nabla^c R $ &  $0$ \\[0.4em]
$R^{ab}  \nabla_c \nabla^c R_{ab} $ &  $-\frac{1}{2} \vec{\d}^6 \hb $ \\[0.4em]
  \hline
 \end{tabular}
 \caption{Contributions to the zero order equations of motion.}
\label{table:EOM_0}
\end{center}
\end{table}

\subsection{First Order Equations of Motion}
In this subsection we detail the rules used to simplify the equations of motion at first order in the probe graviton contribution. We address separately rules which are relevant for the equations of motion of actions with two and three Riemann tensors and those which are relevant for actions with four Riemann tensors.

As explained in subsection \ref{subsec:Gen_con}, on our background and with the assumptions we take regarding the polarisations,  the only components of the higher curvature gravity equations which are not automatically satisfied are $E_{ij}$.
We will use a set of rules to simplify possible structures of $E_{ij}$.

\subsubsection{Actions with Two or Three Riemann Tensors}

Using the statements of the previous subsection and the list of metric components, Christoffel symbols and Riemann components found in appendix \ref{app:List} one can derive the following set of rules for the possible structures of $E_{ij}$. We implemented these rules in \textit{Mathematica} to obtain the results of section  \ref{sec:EOM}.

The vanishing of $R^{(0)}$, $R^{(1)}$ and $R_{\mu\nu}^{(0)}$ implies that up to first order in $h_{ij}$, any contraction of an expression with the following tensor structure should vanish:
\begin{equation}
\begin{split}
 & R=0,
 \qquad
 R_{\alpha\beta} R_{\gamma\delta}=0,
 \qquad
R_{\alpha\beta} \nabla_\gamma \nabla_\delta R_{\epsilon\zeta}=0.
\end{split}
\end{equation}
The fact that there are no zero order quantities with lower $v$  or upper $u$ indices implies that all expressions of the form:
\begin{equation}
\label{rulesR3A}
\begin{split}
 R{}^\alpha {}^\beta {}^\gamma {}^\delta R{}^\rho {}^\sigma {}^\kappa {}^\mu R{}^\nu {}^\zeta {}^\lambda {}^\xi
=0 ,
\end{split}
\end{equation}
with all indices except for two of them  contracted, vanish when they appear in the $ E_{ij}$ component of the equations of motion. This is only to be used in the $R^3$ and $R\nabla^2 R$ actions. The reason is the following:  after extracting all the metric factors, two of the Riemann tensor are at zero order. Each one of them carries two lower $u$ indices. Since the only metric term with upper $u$ indices is $g^{uv}$, this leaves us with four lower $v$ indices. Since no zero order terms have such indices and the first order Riemann tensor can only absorb two of them, such contributions do not arise. The polarisation condition $\epsilon_{ij} b^i b^j = 0$ implies:
\begin{equation}
\begin{split}
 &R{}^\alpha {}^\beta {}^\gamma {}^\delta R{}_\alpha {}_\beta {}_\gamma {}_\delta =0 .
\end{split}
\end{equation}
Using \eqref{chris_van} and \eqref{Riem_van} it is also easy to show that:
\begin{equation}
\label{rulesR3B}
\begin{split}
 &\nabla{}_\alpha R{}_\beta {}_\gamma {}_\delta {}_\rho \nabla{}^\sigma R{}^\mu {}^\nu {}^\zeta {}^\lambda =0, \\
 &\nabla{}_\alpha R{}_\beta {}_\gamma \nabla{}^\sigma R{}^\mu {}^\nu {}^\zeta {}^\lambda =0 ,\\
 &\nabla{}^\alpha R{}^\beta {}^\gamma \nabla{}^\sigma R{}^\mu {}^\nu =0 .
 \end{split}
 \end{equation}
Finally for the non-zero terms we denote:
\begin{equation}
\begin{split}
& {\mathcal{T}_{_{0}}}^{ij}  \equiv  \hb_{,}^{ ~ k j } ~ h^{i  }_{~ k , v v} ,\\
&{\mathcal{T}_{_{1}}}^{ij}  \equiv \ld {\mathcal{T}_{_{0}}}^{ij} ,\\
&\S^{ij}  \equiv \hb_{,}^{ ~ i j k l } ~ h_{ k l , v v},
\end{split}
\end{equation}
where $\ld \equiv -4 ( \d_u \d_v + \hb \d_v ^2 )$,
and obtain the following rules for terms of dimension four:
\begin{equation}
\begin{split}
 &R{}^\alpha {}^\beta R{}^\mu {}_\alpha {}^\nu {}_\beta =0,
 \\
 &R{}^\mu {}^\alpha {}^\beta {}^\gamma R{}^\nu {}_\alpha {}_\beta {}_\gamma
= 2 {\mathcal{T}_{_{0}}} {}^\mu {}^\nu   +2 {\mathcal{T}_{_{0}}} {}^\nu {}^\mu,  \\
 &R{}^\mu {}^\alpha {}^\beta {}^\gamma R{}^\nu {}_\beta {}_\alpha {}_\gamma
=  {\mathcal{T}_{_{0}}} {}^\mu {}^\nu + {\mathcal{T}_{_{0}}} {}^\nu {}^\mu , \\
 &\nabla{}_\alpha \nabla{}^\alpha R{}^\mu {}^\nu = -\frac{1}{2}\ld^2  h{}^\mu {}^\nu  ,
\end{split}
\end{equation}
and of dimension six
\begin{equation}
\begin{split}
 &R{}^\nu {}_\alpha {}_\beta {}_\gamma \nabla{}^\gamma \nabla{}^\beta R{}^\mu {}^\alpha =0, \\
 &R{}^\mu {}_\alpha {}^\nu {}_\beta \nabla{}_\gamma
\nabla{}^\gamma R{}^\alpha {}^\beta =  0  ,\\
 &R{}^\alpha {}^\beta \nabla{}_\gamma
\nabla{}^\gamma R{}^\mu {}_\alpha {}^\nu {}_\beta = 0  ,\\
 &R{}^\alpha {}^\beta \nabla{}_\alpha \nabla{}_\beta R{}^\mu {}^\nu =  0
 ,\\
  &R{}^\nu {}_\alpha {}_\beta {}_\gamma \nabla{}^\gamma
\nabla{}^\mu R{}^\alpha {}^\beta = 0
, \\
 &R{}^\alpha {}^\beta {}^\gamma {}^\delta \nabla{}_\beta
\nabla{}^\mu R{}^\nu {}_\alpha {}_\gamma {}_\delta = -2\, \mathcal{S} {}^\mu {}^\nu  ,\\
 &R{}^\alpha {}^\beta {}^\gamma {}^\delta \nabla{}_\delta
\nabla{}_\beta R{}^\mu {}_\alpha {}^\nu {}_\gamma = \mathcal{S} {}^\mu {}^\nu  ,\\
 &R{}^\nu {}^\alpha {}^\beta {}^\gamma \nabla{}_\rho
\nabla{}^\rho R{}^\mu {}_\alpha {}_\beta {}_\gamma =2\,  {\mathcal{T}_{_{1}}} {}^\mu {}^\nu  ,\\
 &R{}^\nu {}^\gamma {}^\alpha {}^\beta \nabla{}_\gamma \nabla{}_\beta R{}^\mu {}_\alpha =  {\mathcal{T}_{_{1}}} {}^\mu {}^\nu  ,\\
 &R{}^\nu {}^\beta {}^\alpha {}^\gamma \nabla{}_\gamma \nabla{}_\beta R{}^\mu {}_\alpha =  {\mathcal{T}_{_{1}}} {}^\mu {}^\nu  ,\\
 &R{}^\alpha {}^\beta {}^\gamma {}^\delta \nabla{}^\mu
\nabla{}^\nu R{}_\alpha {}_\beta {}_\gamma {}_\delta =4\, \mathcal{S} {}^\mu {}^\nu , \\
 &R{}^\alpha {}^\beta {}^\gamma {}^\delta \nabla{}^\mu
\nabla{}^\nu R{}_\alpha {}_\gamma {}_\beta {}_\delta =2\, \mathcal{S} {}^\mu {}^\nu , \\
 &R{}^\alpha {}^\beta {}^\gamma {}^\delta \nabla{}_\delta
\nabla{}^\mu R{}^\nu {}_\alpha {}_\beta {}_\gamma = \mathcal{S} {}^\mu {}^\nu  ,\\
  &R{}^\nu {}^\alpha {}^\beta {}^\gamma \nabla{}_\delta
\nabla{}^\delta R{}^\mu {}_\beta {}_\alpha {}_\gamma =  {\mathcal{T}_{_{1}}} {}^\mu {}^\nu  ,\\
 &\nabla{}^\alpha \nabla{}_\alpha \nabla{}^\beta \nabla{}_\beta R{}^\mu {}^\nu = -\frac{1}{2} \ld^3   h {}^\mu {}^\nu ,
 \end{split}
 \end{equation}
with $\mu$ and $\nu$ restricted to the transverse directions. We derive the rules for the terms of dimension six as follows. Splitting the covariant derivatives to partial derivatives and Christoffel symbols, we obtain four different kinds of contributions: $R \d^2 R$, $R^2 \d \Gamma$, $R \Gamma \d R$ and $R^2 \Gamma^2$. Due to the relations \eqref{chris_van} and \eqref{Riem_van} the contributions of the form $R \Gamma \d R$ and $R \Gamma^2 R$ vanish at first order in the probe graviton contribution.
The contributions of the form $R^2 \d \Gamma$ are also always vanishing. This however, requires working out explicitly the relevant index structures. The first contribution, of the form $R \d^2 R$, does not vanish and gives the results listed above. We would also like to point out  that terms of the form $g^{ij} \phi$, where $\phi$ is a scalar are automatically vanishing using the same reasoning as in subsection \ref{subsec:Gen_con}.

\subsubsection{Actions with Four Riemann Tensors}
For actions with four Riemann tensors, the equations of motion can contain either four Riemann (or Ricci) tensors, or alternatively three Riemann tensors and two covariant derivatives. Terms in the equations of motion that contain four Riemann tensors have vanishing contribution when evaluated on our background \eqref{hbkd}. This follows from the fact that such contributions would have three $\hb$ factors, accompanied by six lower $u$ indices. To contract them properly one must be able to absorb six lower $v$ indices. However we can absorb at most two lower $v$ indices in the only Riemann tensor which is first order in the probe graviton and two in the free indices of the equation of motion.

For the second possible type of contributions with three Riemann tensors and two covariant derivatives we are only left with structures of the form $ R^2 \del^2 R$. This is derived using  \eqref{rulesR3B} which is still valid  due to the  two beams setup. Focusing on contributions to the transverse equations of motion $E_{ij}$, we have two of the Riemann tensors that have to be of zero order in the probe graviton contributions and are therefore associated with four lower $u$ indices. We are left with four lower $v$ indices that need to be absorbed. Two of them can be absorbed in a first order term and the two others in the derivatives. As such, the derivatives cannot have free indices and have to be of the form $\del_v$.

Requiring two $v$ indices to be absorbed by a first order term forces it to be a Riemann tensor (since the Ricci tensor cannot absorb such $v$ indices). We therefore have:
\begin{align}
\begin{split}
  &  R{}^\alpha {}^\beta {}^\gamma {}^\delta R{}^\rho {}^\sigma {}^\kappa {}^\mu \nabla{}^\zeta
\nabla{}^\lambda R{}^\nu {}^\xi = 0, \\
 &
 R{}^\alpha {}^\beta R{}^\rho {}^\sigma {}^\kappa {}^\mu \nabla{}^\zeta \nabla{}^\lambda R{}^\nu {}^\xi =
0, \\
 &R{}^\alpha {}^\beta R{}^\rho {}^\sigma \nabla{}^\zeta \nabla{}^\lambda R{}^\nu {}^\xi =
0, \\
 &R{}^\alpha {}^\beta R{}^\rho {}^\sigma \nabla{}^\zeta \nabla{}^\lambda R{}^\gamma {}^\delta {}^\nu {}^\xi =
0, \\
 &R{}^\alpha {}^\beta R{}^\rho {}^\sigma {}^\kappa {}^\mu \nabla{}^\zeta
\nabla{}^\lambda R{}^\nu {}^\xi {}^\gamma {}^\delta = 0.
\end{split}
\end{align}
The requirement the derivatives must absorb two lower $v$ indices automatically forbids terms of the form $R R \nabla_\alpha \nabla^\alpha R$, $ R R \nabla \nabla^\mu R$ and  $ R R \nabla^\mu \nabla^\nu R$ where $\mu$ and $\nu$ are the free transverse indices of the equation of motion.

We also notice that the two Riemann tensors outside of the derivatives must be at zeroth order. This is because the derivative of a zeroth order quantity with respect to $v$ is vanishing, which forces the Riemann inside the derivatives to be first order. Hence we cannot contract indices which are not transverse between the two Riemann tensors outside the derivatives. Remembering that the free indices are transverse we obtain:
\begin{align}
\begin{split}
 &R{}_\alpha {}^\delta {}^\rho {}^\sigma R{}^\nu {}^\alpha {}^\beta {}^\gamma \nabla{}_\sigma
\nabla{}_\gamma R{}^\mu {}_\delta {}_\beta {}_\rho =0, \\
&
R{}_\alpha {}^\delta {}^\rho {}^\sigma R{}^\nu {}^\alpha {}^\beta {}^\gamma \nabla{}_\gamma
\nabla{}_\beta R{}^\mu {}_\delta {}_\rho {}_\sigma =0, \\
 &R{}_\alpha {}^\delta {}^\rho {}^\sigma R{}^\nu {}^\alpha {}^\beta {}^\gamma \nabla{}_\delta
\nabla{}_\gamma R{}^\mu {}_\beta {}_\rho {}_\sigma =0, \\
 &R{}_\alpha {}^\delta {}^\rho {}^\sigma R{}^\nu {}^\alpha {}^\beta {}^\gamma \nabla{}_\sigma
\nabla{}_\delta R{}^\mu {}_\rho {}_\beta {}_\gamma =0, \\
&R{}^\nu {}^\alpha {}^\beta {}^\gamma R{}_\alpha {}^\delta {}^\rho {}^\sigma \nabla{}_\sigma
\nabla{}_\gamma R{}^\mu {}_\rho {}_\beta {}_\delta =0 .
\end{split}
\end{align}
Direct calculation then leads to
\begin{align}
\begin{split}
 &R{}_\beta {}^\delta {}^\rho {}^\sigma R{}^\nu {}^\alpha {}^\beta {}^\gamma \nabla{}_\sigma
\nabla{}_\gamma R{}^\mu {}_\rho {}_\alpha {}_\delta = 0, \\
 &
  R{}_\beta {}^\delta {}^\rho {}^\sigma R{}^\nu {}^\alpha {}^\beta {}^\gamma \nabla{}_\sigma
\nabla{}_\alpha R{}^\mu {}_\rho {}_\gamma {}_\delta= 0 ,
\end{split}
 \\
 \begin{split}
 &R{}^\mu {}^\alpha {}^\beta {}^\gamma R{}^\nu {}^\delta {}^\rho {}^\sigma \nabla{}_\gamma
\nabla{}_\beta R{}_\alpha {}_\delta {}_\rho {}_\sigma =0, \\
 &
R{}^\nu {}^\alpha {}^\beta {}^\gamma R{}_\alpha {}^\delta {}^\rho {}^\sigma \nabla{}_\gamma
\nabla{}_\beta R{}^\mu {}_\delta {}_\rho {}_\sigma= 0 ,
\end{split}
\\
\begin{split}
 &R{}^\mu {}^\alpha {}^\nu {}^\beta R{}^\gamma {}^\delta {}^\rho {}^\sigma \nabla{}_\delta
\nabla{}_\alpha R{}_\beta {}_\gamma {}_\rho {}_\sigma =0 , \\
 &
 R{}^\mu {}^\alpha {}^\nu {}^\beta R{}^\gamma {}^\delta {}^\rho {}^\sigma \nabla{}_\delta
\nabla{}_\beta R{}_\alpha {}_\gamma {}_\rho {}_\sigma= 0 ,
\end{split}
\\
\begin{split}
&R{}^\mu {}^\alpha {}^\nu {}^\beta  R{}^\gamma {}^\delta {}^\rho {}^\sigma \nabla{}_\beta
\nabla{}_\alpha R{}_\gamma {}_\delta {}_\rho {}_\sigma =0 , \\
 &
  R{}^\mu {}^\alpha {}^\nu {}^\beta R{}^\gamma {}^\delta {}^\rho {}^\sigma \nabla{}_\sigma
\nabla{}_\delta R{}_\alpha {}_\gamma {}_\beta {}_\rho =0 ,
\end{split}
\\
 &R{}^\alpha {}^\beta {}^\gamma {}^\delta R{}_\alpha {}_\beta {}^\rho {}^\sigma \nabla{}_\sigma
\nabla{}_\delta R{}^\mu {}_\gamma {}^\nu {}_\rho =0, \\
 & R{}^\alpha {}^\beta {}^\gamma {}^\delta R{}_\alpha {}^\rho {}^\sigma {}^\xi \nabla{}_\xi
\nabla{}_\rho R{}_\beta {}_\gamma {}_\delta {}_\sigma= 0  ,
 \end{align}
Finally, defining:
\begin{align}
\begin{split}
 \Uz {}^i {}^j = \hb_{,}^{~lk} \hb_{, k}{}^{ j} h^{i }_{~~ l , v^4} \ ,
\\
\Wz{}^i {}^j =   \hb_{,}^{~ik} \hb_{,}^{~jl} h_{kl , v^4} \ ,
\\
 \Vz{}^i {}^j = \hb_{,}^{~kl} \hb_{,kl} h^{ij}_{~~ , v^4} \ ,
\end{split}
\end{align}
the non-vanishing contributions are given by:
\begin{align}
\begin{split}
  &R{}_\alpha {}_\beta {}_\gamma {}_\delta R{}^\rho {}^\beta {}^\sigma
{}^\delta \nabla{}_\sigma \nabla{}^\gamma R{}^\mu {}^\alpha {}^\nu {}_\rho = -2 \Vz{}^\mu {}^\nu, \\
 &
 R{}_\alpha {}^\rho {}_\gamma {}^\sigma R{}^\alpha {}^\beta {}^\gamma {}^\delta \nabla{}_\sigma
\nabla{}_\rho R{}^\mu {}_\beta {}^\nu {}_\delta =-2 \Vz{}^\mu {}^\nu, \\
 &R{}_\alpha {}^\rho {}_\gamma {}^\sigma R{}^\alpha {}^\beta {}^\gamma {}^\delta \nabla{}_\sigma
\nabla{}_\delta R{}^\mu {}_\beta {}^\nu {}_\rho =-2  \Vz{}^\mu {}^\nu , \\
 &R{}_\alpha {}^\rho {}_\gamma {}^\sigma R{}^\alpha {}^\beta {}^\gamma {}^\delta \nabla{}_\rho
\nabla{}_\delta R{}^\mu {}_\beta {}^\nu {}_\sigma =-2  \Vz{}^\mu {}^\nu,
\end{split}
\\
 &R{}_\alpha {}_\beta {}_\gamma {}_\delta R{}^\nu {}^\sigma {}^\alpha {}^\rho \nabla{}_\sigma
\nabla{}_\rho R{}^\mu {}^\beta {}^\gamma {}^\delta =-4\Uz{}^\mu {}^\nu  ,
\\
\begin{split}
 &R{}_\beta {}^\delta {}^\rho {}^\sigma R{}^\nu {}^\alpha {}^\beta {}^\gamma \nabla{}_\sigma
\nabla{}_\gamma R{}^\mu {}_\delta {}_\alpha {}_\rho =2\Uz{}^\mu {}^\nu , \\
 &
 R{}_\beta {}^\delta {}^\rho {}^\sigma R{}^\nu {}^\alpha {}^\beta {}^\gamma \nabla{}_\sigma
\nabla{}_\delta R{}^\mu {}_\alpha {}_\gamma {}_\rho =2\Uz{}^\mu {}^\nu , \\
 &R{}_\beta {}^\delta {}^\rho {}^\sigma R{}^\nu {}^\alpha {}^\beta {}^\gamma \nabla{}_\sigma
\nabla{}_\alpha R{}^\mu {}_\delta {}_\gamma {}_\rho =2\Uz{}^\mu {}^\nu  , \\
 &R{}_\beta {}^\delta {}^\rho {}^\sigma R{}^\nu {}^\alpha {}^\beta {}^\gamma \nabla{}_\sigma
\nabla{}_\delta R{}^\mu {}_\gamma {}_\alpha {}_\rho =2\Uz{}^\mu {}^\nu  ,
\end{split}
\\
\begin{split}
 &R{}_\beta {}^\delta {}^\rho {}^\sigma R{}^\nu {}^\alpha {}^\beta {}^\gamma \nabla{}_\gamma
\nabla{}_\alpha R{}^\mu {}_\delta {}_\rho {}_\sigma =-4\Uz{}^\mu {}^\nu  , \\
 &
 R{}_\beta {}^\delta {}^\rho {}^\sigma R{}^\nu {}^\alpha {}^\beta {}^\gamma \nabla{}_\delta
\nabla{}_\gamma R{}^\mu {}_\alpha {}_\rho {}_\sigma =-4\Uz{}^\mu {}^\nu , \\
 &R{}_\beta {}^\delta {}^\rho {}^\sigma R{}^\nu {}^\alpha {}^\beta {}^\gamma \nabla{}_\delta
\nabla{}_\alpha R{}^\mu {}_\gamma {}_\rho {}_\sigma =-4\Uz{}^\mu {}^\nu  ,
\end{split}
\\
 &R{}^\mu {}^\alpha {}^\beta {}^\gamma R{}^\nu {}^\delta {}^\rho {}^\sigma \nabla{}_\gamma
\nabla{}_\alpha R{}_\beta {}_\delta {}_\rho {}_\sigma =-4\Wz{}^\mu {}^\nu , \\
 &R{}^\mu {}^\alpha {}^\beta {}^\gamma R{}^\nu {}^\delta {}^\rho {}^\sigma \nabla{}_\sigma
\nabla{}_\gamma R{}_\alpha {}_\rho {}_\beta {}_\delta =2\Wz{}^\mu {}^\nu  , \\
 &R{}^\mu {}^\alpha {}^\beta {}^\gamma R{}^\nu {}^\delta {}^\rho {}^\sigma \nabla{}_\delta
\nabla{}_\alpha R{}_\beta {}_\gamma {}_\rho {}_\sigma =-8\Wz{}^\mu {}^\nu ,
\end{align}
where some of the rules are related using the second Bianchi identity.

\bibliographystyle{utphys}
\bibliography{GravBkg}

\providecommand{\href}[2]{#2}\begingroup\raggedright\begin{thebibliography}{10}

\bibitem{camanho_causality_2014}
X.~O. Camanho, J.~D. Edelstein, J.~Maldacena, and A.~Zhiboedov, ``{Causality
  Constraints on Corrections to the Graviton Three-Point Coupling},''
\href{http://arxiv.org/abs/1407.5597}{{\ttfamily arXiv:1407.5597 [hep-th]}}.

\bibitem{dray_gravitational_1985}
T.~Dray and G.~'t~Hooft, ``The {Gravitational} {Shock} {Wave} of a {Massless}
  {Particle},'' \href{http://dx.doi.org/10.1016/0550-3213(85)90525-5}{{\em
  Nucl.Phys.} {\bfseries B253} (1985) 173--188}.

\bibitem{dappollonio_regge_2015}
G.~D'Appollonio, P.~Vecchia, R.~Russo, and G.~Veneziano, ``{Regge behavior
  saves String Theory from causality violations},''
  \href{http://dx.doi.org/10.1007/JHEP05(2015)144}{{\em JHEP} {\bfseries 05}
  (2015) 144},
\href{http://arxiv.org/abs/1502.01254}{{\ttfamily arXiv:1502.01254 [hep-th]}}.

\bibitem{papallo_graviton_2015}
G.~Papallo and H.~S. Reall, ``{Graviton time delay and a speed limit for small
  black holes in Einstein-Gauss-Bonnet theory},''
  \href{http://dx.doi.org/10.1007/JHEP11(2015)109}{{\em JHEP} {\bfseries 11}
  (2015) 109},
\href{http://arxiv.org/abs/1508.05303}{{\ttfamily arXiv:1508.05303 [gr-qc]}}.

\bibitem{Deser:2014hga}
S.~Deser, M.~Sandora, A.~Waldron, and G.~Zahariade, ``{Covariant constraints
  for generic massive gravity and analysis of its characteristics},''
  \href{http://dx.doi.org/10.1103/PhysRevD.90.104043}{{\em Phys. Rev.}
  {\bfseries D90} no.~10, (2014) 104043},
\href{http://arxiv.org/abs/1408.0561}{{\ttfamily arXiv:1408.0561 [hep-th]}}.

\bibitem{Deser:2014fta}
S.~Deser, K.~Izumi, Y.~C. Ong, and A.~Waldron, ``{Problems of massive
  gravities},'' \href{http://dx.doi.org/10.1142/S0217732315400064}{{\em Mod.
  Phys. Lett.} {\bfseries A30} (2015) 1540006},
\href{http://arxiv.org/abs/1410.2289}{{\ttfamily arXiv:1410.2289 [hep-th]}}.

\bibitem{Sarbach:2012pr}
O.~Sarbach and M.~Tiglio, ``{Continuum and Discrete Initial-Boundary-Value
  Problems and Einstein's Field Equations},''
  \href{http://dx.doi.org/10.12942/lrr-2012-9}{{\em Living Rev. Rel.}
  {\bfseries 15} (2012) 9},
\href{http://arxiv.org/abs/1203.6443}{{\ttfamily arXiv:1203.6443 [gr-qc]}}.

\bibitem{courant_methods_1962}
R.~Courant and D.~Hilbert, {\em Methods of {Mathematical} {Physics}}, vol.~2.
\newblock Wiley, New York, 1962.

\bibitem{Hormander:1983}
L.~H{\"o}rmander, {\em The analysis of linear partial differential operators,
  Vol I and II}.
\newblock Classics in Mathematics. Springer-Verlag, Berlin, 1983.
\newblock Distribution theory and Fourier analysis.

\bibitem{Gao:2000ga}
S.~Gao and R.~M. Wald, ``{Theorems on gravitational time delay and related
  issues},'' \href{http://dx.doi.org/10.1088/0264-9381/17/24/305}{{\em Class.
  Quant. Grav.} {\bfseries 17} (2000) 4999--5008},
\href{http://arxiv.org/abs/gr-qc/0007021}{{\ttfamily arXiv:gr-qc/0007021
  [gr-qc]}}.

\bibitem{Visser:1998ua}
M.~Visser, B.~Bassett, and S.~Liberati, ``{Superluminal censorship},''
  \href{http://dx.doi.org/10.1016/S0920-5632(00)00782-9}{{\em Nucl. Phys. Proc.
  Suppl.} {\bfseries 88} (2000) 267--270},
\href{http://arxiv.org/abs/gr-qc/9810026}{{\ttfamily arXiv:gr-qc/9810026
  [gr-qc]}}.

\bibitem{Palmer:2002vn}
B.~C. Palmer and D.~Marolf, ``{On fast travel through spherically symmetric
  space-times},'' \href{http://dx.doi.org/10.1103/PhysRevD.67.044012}{{\em
  Phys. Rev.} {\bfseries D67} (2003) 044012},
\href{http://arxiv.org/abs/gr-qc/0211045}{{\ttfamily arXiv:gr-qc/0211045
  [gr-qc]}}.

\bibitem{Olum:1998mu}
K.~D. Olum, ``{Superluminal travel requires negative energies},''
  \href{http://dx.doi.org/10.1103/PhysRevLett.81.3567}{{\em Phys. Rev. Lett.}
  {\bfseries 81} (1998) 3567--3570},
\href{http://arxiv.org/abs/gr-qc/9805003}{{\ttfamily arXiv:gr-qc/9805003
  [gr-qc]}}.

\bibitem{Woodard:2006nt}
R.~P. Woodard, ``{Avoiding dark energy with 1/r modifications of gravity},''
  \href{http://dx.doi.org/10.1007/978-3-540-71013-4_14}{{\em Lect. Notes Phys.}
  {\bfseries 720} (2007) 403--433},
\href{http://arxiv.org/abs/astro-ph/0601672}{{\ttfamily arXiv:astro-ph/0601672
  [astro-ph]}}.

\bibitem{Chen:2012au}
T.-j. Chen, M.~Fasiello, E.~A. Lim, and A.~J. Tolley, ``{Higher derivative
  theories with constraints: Exorcising Ostrogradski's Ghost},''
  \href{http://dx.doi.org/10.1088/1475-7516/2013/02/042}{{\em JCAP} {\bfseries
  1302} (2013) 042},
\href{http://arxiv.org/abs/1209.0583}{{\ttfamily arXiv:1209.0583 [hep-th]}}.

\bibitem{drummond_qed_1980}
I.~T. Drummond and S.~J. Hathrell, ``{QED} {Vacuum} {Polarization} in a
  {Background} {Gravitational} {Field} and {Its} {Effect} on the {Velocity} of
  {Photons},'' \href{http://dx.doi.org/10.1103/PhysRevD.22.343}{{\em Phys.Rev.}
  {\bfseries D22} (1980) 343}.

\bibitem{Hollowood:2015elj}
T.~J. Hollowood and G.~M. Shore, ``{Causality Violation, Gravitational
  Shockwaves and UV Completion},''
\href{http://arxiv.org/abs/1512.04952}{{\ttfamily arXiv:1512.04952 [hep-th]}}.

\bibitem{bonnor1969gravitational}
W.~Bonnor, ``The gravitational field of light,'' {\em Communications in
  Mathematical Physics} {\bfseries 13} no.~3, (1969) 163--174.

\bibitem{Goroff:1983hc}
M.~Goroff and J.~H. Schwarz, ``{$D$-dimensional Gravity in the Light Cone
  Gauge},''
\href{http://dx.doi.org/10.1016/0370-2693(83)91630-1}{{\em Phys. Lett.}
  {\bfseries B127} (1983) 61--64}.

\bibitem{Babichev:2007dw}
E.~Babichev, V.~Mukhanov, and A.~Vikman, ``{k-Essence, superluminal
  propagation, causality and emergent geometry},''
  \href{http://dx.doi.org/10.1088/1126-6708/2008/02/101}{{\em JHEP} {\bfseries
  02} (2008) 101},
\href{http://arxiv.org/abs/0708.0561}{{\ttfamily arXiv:0708.0561 [hep-th]}}.

\bibitem{reall_causality_2014}
H.~Reall, N.~Tanahashi, and B.~Way, ``Causality and {Hyperbolicity} of
  {Lovelock} {Theories},''
  \href{http://dx.doi.org/10.1088/0264-9381/31/20/205005}{{\em
  Class.Quant.Grav.} {\bfseries 31} (2014) 205005}.

\bibitem{martin-garcia_invar_2007}
J.~M. Mart\'in-Garc\'ia, R.~Portugal, and L.~R.~U. Manssur, ``The {Invar}
  {Tensor} {Package},'' \href{http://dx.doi.org/10.1016/j.cpc.2007.05.015}{{\em
  Comput.Phys.Commun.} {\bfseries 177} (2007) 640--648}.

\bibitem{martin-garcia_invar_2008}
J.~M. Mart\'in-Garc\'ia, D.~Yllanes, and R.~Portugal, ``The {Invar} tensor
  package: {Differential} invariants of {Riemann},''
  \href{http://dx.doi.org/10.1016/j.cpc.2008.04.018}{{\em Comput.Phys.Commun.}
  {\bfseries 179} (2008) 586--590}.

\bibitem{martin-garcia_xperm:_2008}
J.~M. Mart\'in-Garc\'ia, ``{xPerm}: fast index canonicalization for tensor
  computer algebra,'' {\em Computer physics communications} {\bfseries 179}
  no.~8, (2008) 597--603.

\bibitem{nutma_xtras:_2014}
T.~Nutma, ``{xTras}: {A} field-theory inspired {xAct} package for
  mathematica,'' \href{http://dx.doi.org/10.1016/j.cpc.2014.02.006}{{\em
  Comput.Phys.Commun.} {\bfseries 185} (June, 2014) 1719--1738}.

\bibitem{tseytlin_r**4_2000}
A.~A. Tseytlin, ``R**4 terms in 11 dimensions and conformal anomaly of (2,0)
  theory,'' \href{http://dx.doi.org/10.1016/S0550-3213(00)00380-1}{{\em
  Nucl.Phys.} {\bfseries B584} (2000) 233--250}.

\bibitem{gross_superstring_1986}
D.~J. Gross and E.~Witten, ``Superstring {Modifications} of {Einstein}'s
  {Equations},'' \href{http://dx.doi.org/10.1016/0550-3213(86)90429-3}{{\em
  Nucl.Phys.} {\bfseries B277} (1986) 1}.

\bibitem{bellazzini_quantum_2015}
B.~Bellazzini, C.~Cheung, and G.~N. Remmen, ``{Quantum Gravity Constraints from
  Unitarity and Analyticity},'' {\em Phys. Rev.} {\bfseries D92} (2015) 125009,
\href{http://arxiv.org/abs/1509.00851}{{\ttfamily arXiv:1509.00851 [hep-th]}}.

\end{thebibliography}\endgroup

\end{document}